\documentclass[%
twocolumn,
superscriptaddress,
groupedaddress,
showpacs,
 aps,
 prb,
10pt
]{revtex4-2}

\usepackage{mathtools}
\usepackage{amsmath}
\usepackage{amssymb}
\usepackage{bbm}
\usepackage{bm}
\usepackage{graphicx}
\usepackage{xcolor}
\usepackage{dcolumn}
\usepackage{bm}
\usepackage{float}
\usepackage{subfigure}
\usepackage[colorlinks]{hyperref}
\usepackage{dsfont}
\usepackage{braket}
\usepackage{units}
\usepackage{placeins}

\newcommand{\mat}[1]{\bm{#1}}
\newcommand{\matK}[1]{\underline{#1}} 
\renewcommand{\vec}[1]{\bm{#1}}

\DeclareMathOperator*{\IIm}{Im}
\newcommand{\ee}{\mathrm{e}}  
\newcommand{\ii}{\mathrm{i}} 
\DeclareMathOperator*{\Tr}{Tr} 
\DeclareMathOperator*{\dd}{d} 

\begin{document}


\title{Configuration interaction based nonequilibrium steady state impurity solver}

\author{Daniel Werner}
\email[]{daniel.werner96@posteo.at}
\affiliation{Institute of Theoretical and Computational Physics, Graz University of Technology, 8010 Graz, Austria}
\author{Jan Lotze}
\affiliation{Institute of Theoretical and Computational Physics, Graz University of Technology, 8010 Graz, Austria}
\author{Enrico Arrigoni}
\email[]{arrigoni@tugraz.at}
\affiliation{Institute of Theoretical and Computational Physics, Graz University of Technology, 8010 Graz, Austria}

\date{\today}

\begin{abstract}
We present a solver for correlated impurity problems out of equilibrium based on a combination of the so-called auxiliary master equation approach (AMEA) and the configuration interaction expansion. Within AMEA one maps the original impurity model onto an auxiliary open quantum system with a restricted number of bath sites which can be addressed by numerical many-body approaches such as Lanczos/Arnoldi exact diagonalization (ED) or matrix product states (MPS). While the mapping becomes exponentially more accurate with increasing number of bath sites, ED implementations are severely limited due to the fast increase of the Hilbert space dimension for open systems, and the MPS solver typically requires rather long runtimes. 
Here, we propose to adopt a configuration interaction approach augmented by active space extension to solve numerically the correlated auxiliary open quantum system. This allows access to a larger number of bath sites at lower computational costs than for plain ED. We benchmark the approach with numerical renormalization group results in equilibrium and with MPS out of equilibrium. In particular, we evaluate the current, the conductance as well as the Kondo peak and its splitting as a function of increasing bias voltage  below the Kondo temperature $T_{\text{K}}$. We obtain a rather accurate scaling of the conductance as a function of the  bias voltage and temperature rescaled by $T_{\text{K}}$ for moderate to strong interactions in a wide range of parameters. The approach combines the fast runtime of ED with an accuracy close to the one achieved by MPS making it an attractive  solver for nonequilibrium dynamical mean field theory.
\end{abstract}


\maketitle

\section{\label{sec:intro}Introduction}
The single-impurity Anderson model~\cite{ande.61, kond.64, sc.wo.66}, was originally devised to gain insight into the effect of dilute magnetic impurities in metals. Its universal low-energy physics characterized by the Kondo temperature $T_{\text{K}}$ as energy scale can be captured quantitatively by the numerical renormalization group (NRG)~\cite{wils.75, bu.co.08, zi.pr.09}. Besides the description of quantum dots~\cite{go.sh.98,wi.fr.00,fr.ha.02,le.sc.05}, nanowires~\cite{ra.bu.94,kr.sh.11,kr.sh.12} and carbon nanotubes~\cite{fe.ar.17}, an important application is the dynamical mean field theory (DMFT)~\cite{me.vo.89, ge.ko.92, ge.ko.96}, where a single site in a correlated lattice is identified with an impurity embedded in a self-consistently determined effective environment. Using the Keldysh formalism, DMFT can be extended to the nonequilibrium case~\cite{fr.tu.06, sc.mo.02u, ao.ts.14}. 

In this paper we will focus on a correlated impurity out of equilibrium and investigate its properties with an implementation of the so-called auxiliary master equation approach (AMEA)~\cite{ar.kn.13,do.nu.14} based on the configuration interaction (CI) expansion and its complete active space (CASCI) extension~\cite{he.jo, sh.sc.99, zg.gu.12, li.de.13}. AMEA is used to study nonequilibrium steady state properties of strongly correlated impurity models, especially in connection with DMFT~\cite{ti.do.15,ti.do.16,ti.so.18}. It is based on a mapping of the original impurity plus bath model onto an auxiliary open quantum system, which becomes exponentially exact upon increasing the number of its bath sites~\cite{do.so.17}. This auxiliary many-body impurity problem has been solved by either Lanczos/Arnoldi exact diagonalization (ED)~\cite{ar.kn.13,do.nu.14}, matrix product states (MPS)~\cite{do.ga.15,fu.do.18}, or stochastic wave-function approaches~\cite{so.fu.19}. 

While the last two schemes provide more accurate results for the impurity problem itself, they are numerically quite expensive. Hence for DMFT, where a large number of impurity solutions are required to achieve self-consistency, only the ED solver has been adopted so far. Unfortunately, since the dimension of the many-body density matrix of the open quantum systems grows as the square of the dimension of the Hilbert space, the number of bath sites is limited to a maximum of $N_{\text{B}}=6$ for ED. On the other hand, the corresponding Lindblad equation provides a larger number of parameters to fit the bath hybridization function so that roughly speaking the same accuracy can be achieved as for a closed system with twice as many bath sites in equilibrium~\cite{do.so.17}. Issues arise when the hybridization function becomes too sharp or exhibits too many peaks which cannot be captured by the parameters provided by six bath sites. The first situation can occur when investigating the Kondo peak for high Hubbard interactions at low temperature, while the second one can become relevant when performing DMFT.

Due to the exponential increase of the accuracy with the number of bath sites, it would be desirable to access even a few more of them without a significant increase of the numerical costs. In this paper we show that these requirements are met by CI (and its extension CASCI), which have been shown to be quite accurate as many-body solvers for equilibrium (closed) impurity models for which the interaction is localized on one or a few sites~\cite{zg.gu.12,zg.ch.11}. In our case, these approaches allow to consider up to $N_{\text{B}}=10$ bath sites while the runtime does not significantly increase beyond that of the ED solver for up to $N_{\text{B}}=8$. To demonstrate their capabilities, we benchmark CI and CASCI against MPS by comparing the spectral function of the Anderson impurity model (AIM) in and out of equilibrium. As a particularly challenging benchmark, we address the equilibrium problem, which is accessible by AMEA as well, while certainly being less accurate than more established methods such as NRG. Here, we compare the temperature dependence of the equilibrium conductance obtained with CI and CASCI against NRG results down below the Kondo temperature. Thereby, the computed conductance quite accurately collapses into a single function of the scaled temperature $T/T_{\text{K}}$ for different moderate to strong interactions and for $T \lesssim 0.1 D$, where $D$ is the half-bandwidth. This scaling behavior is also quite accurately reproduced  when considering the nonequilibrium conductance as a function  of the scaled bias voltage $\phi/T_{\text{K}}$ for the same $T/T_{\text{K}}$.  Finally, we evaluate the nonequilibrium spectral function as function of the bias voltage and comment on the splitting of the Kondo peak.

One should mention that recently very accurate results for the nonequilibrium steady state spectral function of the AIM have been obtained by inchworm Quantum Monte Carlo (QMC)~\cite{er.gu.22u}, which is also a numerically quite expensive approach. Among other approaches to deal with time-dependent impurity problems up to long times one should mention quantum quasi Monte Carlo (see, e.g.~\cite{be.ba.21}) and fork tensor network~\cite{ba.zi.17} approaches. Also numerical and perturbative renormalization-group approaches~\cite{wils.75,me.ki.13,ha.we.14,bu.co.08,zi.pr.09} obviously achieve much more accurate scaling behaviors. The strength of our approach, besides being applicable for strong and weak interactions, lies in the considerably reduced computational costs, especially the wall time, making it attractive as a nonequilibrium impurity solver for DMFT

The paper is organized as follows. In Sec.~\ref{sec:methods}, we review AMEA, before discussing CI in Sec.~\ref{sec:CI}. Starting in Sec.~\ref{sec:results}, we determine the method parameter of CI in Sec.~\ref{sec:dof}. Then, in Sec.~\ref{sec:real_results}, we benchmark it against the NRG~\cite{zitk.21} conductance in equilibrium and MPS spectral functions out of equilibrium~\cite{do.ga.15}. We conclude in Sec.~\ref{sec:conclusion}. The description of and results obtained using CASCI are provided in App.~\ref{sec:CASCI}.

\section{\label{sec:methods}Model and Method}
\subsection{\label{sec:KGF}Nonequilibrium Green's functions}
A convenient way to represent Green's functions in the nonequilibrium case is via the Keldysh formalism~\cite{schw.61, kad.baym, keld.65, ra.sm.86, ha.ja, wagn.91}. Since we are interested in the steady state where time translational invariance applies, we work in the frequency domain
\begin{equation}\label{eqn:GF_full}
\matK{G}(\omega) = 
\left(\begin{matrix}
G^\text{R}(\omega) & G^\text{K}(\omega) \\
0 & G^\text{A}(\omega)
\end{matrix} \right),
\end{equation}
where $\matK{G}(\omega)$ is used to denote the $2\times 2$ matrix structure and $G^\text{A}(\omega) = [G^\text{R}(\omega)]^\dagger$. 

In equilibrium, the Keldysh component of the Green's function can be determined from the retarded one using the fluctuation dissipation theorem
\begin{equation}\label{eqn:fluc_diss}
G_{}^\text{K}(\omega) = 2\ii[1-2\rho_{\text{FD}}^{}(\omega, \mu, T)]\IIm[G_{}^\text{R}(\omega)],
\end{equation}
where $\rho_{\text{FD}}^{}(\omega, \mu, T)=\{\exp[(\omega-\mu)/T]+1\}^{-1}$ is the Fermi-Dirac distribution function. 

The spectral function (DOS) is given by
\begin{equation}\label{eqn:spectral}
A(\omega) = -\frac{1}{\pi} \IIm[G_{}^\text{R}(\omega)].
\end{equation}

\subsection{\label{sec:imp_model}Physical impurity model}
\begin{figure}[t]
\centering
\includegraphics[width=\linewidth]{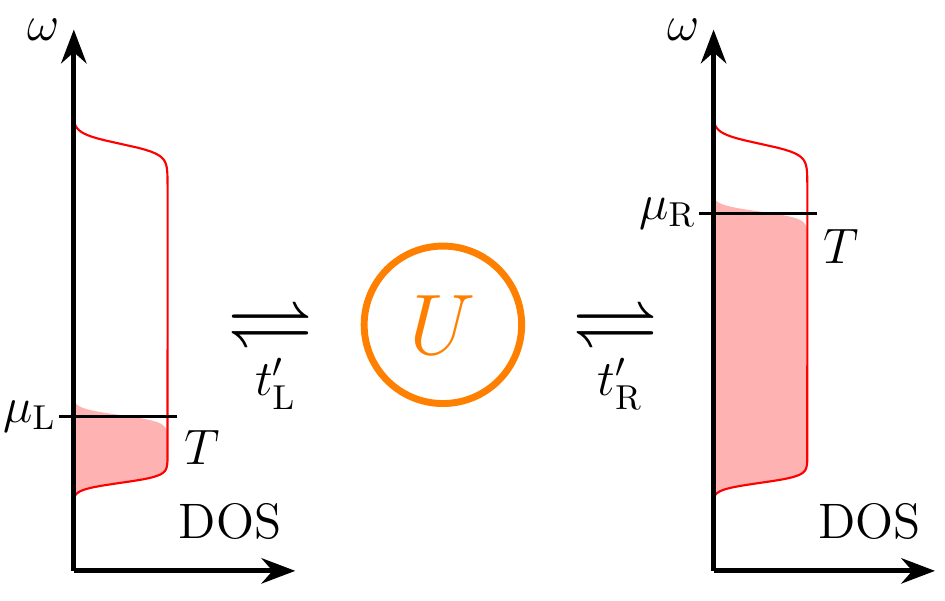}
\caption{Physical system consisting of an impurity with Hubbard interaction $U$ coupled via $t_{\text{L,R}}^{\prime}$ to two leads with flat band spectrum characterized by the chemical potentials $\mu_{\text{L,R}}$ and temperature $T$.}
\label{fig:physical_system}
\end{figure}

The physical system consists of a correlated impurity coupled to two noninteracting leads, as sketched in Fig.~\ref{fig:physical_system} and described by the Hamiltonian
\begin{equation}\label{eqn:H_full}
H = H_{\text{imp}}^{} + H_{\text{bath}}^{} + H_{\text{coup}}^{}.
\end{equation}
Here $H_{\text{imp}}$ denotes the impurity Hamiltonian
\begin{equation}\label{eqn:H_imp}
H_{\text{imp}}^{} = \varepsilon_{\text{imp}}^{} \sum_\sigma d_{\sigma}^\dagger d_{\sigma}^{} + U n_{\text{d}\uparrow}^{} n_{\text{d}\downarrow}^{}
\end{equation}
with Hubbard interaction $U$, on-site energy $\varepsilon_{\text{imp}}^{}$, creation (annihilation) operator $d_{\sigma}^{\dagger}$ ($d_{\sigma}^{}$) of a fermion at the impurity site with spin $\sigma$ and particle number operator $n_{\text{d}\sigma}=d_{\sigma}^{\dagger}d_{\sigma}^{}$. $H_{\text{bath}}^{}$ describes the leads
\begin{equation}\label{eqn:H_res}
H_{\text{bath}}^{} = \sum_{k \lambda \sigma}^{} \varepsilon_{\lambda k}^{} a_{\lambda k \sigma}^\dagger a_{\lambda k \sigma}^{},
\end{equation}
with dispersion $\varepsilon_{\lambda k}^{}$ and creation (annihilation) operator $a_{\lambda k \sigma}^{\dagger}$ ($a_{\lambda k \sigma}^{}$) of a fermion in the left and right lead $\lambda \in \{\text{L},\text{R}\}$ labeled by momentum $k$. The coupling between impurity and bath is given by
\begin{equation}\label{eqn:H_coup}
H_{\text{coup}}^{} = \frac{1}{\sqrt{N_k^{}}}\sum_{k \lambda \sigma} t_\lambda^{\prime} (a_{\lambda k \sigma}^\dagger d_{\sigma}^{} + d_{\sigma}^\dagger a_{\lambda k \sigma}^{}),
\end{equation}
where $t_\lambda^{\prime}$ is the coupling strength between the leads and the impurity, and $N_{k}^{}\to\infty$ is the number of $k$ points. In this paper we only consider the particle hole symmetric case with $\varepsilon_{\text{imp}}^{} = -U/2$, $t_\text{L}^{\prime} = t_\text{R}^{\prime}$, and $\varepsilon_{\text{L} k}^{} = \varepsilon_{\text{R} k}^{}$.

The environment of the impurity can be captured by the physical hybridization function, which one can write~\cite{economou} as
\begin{equation}\label{eqn:Delta_ph}
\matK{\Delta}_{\text{ph}}^{} = \sum_\lambda t_{\lambda}^{\prime 2} \matK{g}_\lambda^{}(\omega),
\end{equation}
where $\matK{g}_\lambda^{}(\omega)$ are the Green's functions of the leads at the point of contact. In the following we consider leads with a flat band spectrum
\begin{equation}\label{eqn:hyb_ph}
-\IIm[g_\lambda^{\text{R}}(\omega)] = \frac{\pi}{2D}\Theta(D-|\omega|).
\end{equation}
For convenience, we choose $t_\lambda^{\prime} = \sqrt{\Gamma D/\pi}$ giving $-\IIm[\Delta_{\text{ph}}^{}(0)] = \Gamma$, which is the unit of energy throughout this paper. This spectrum is used for the reference calculations with NRG. Since the precise form of the cutoff is unimportant, for AMEA where the hybridization function is fitted, it is convenient to employ a smoothed version of Eq.~\eqref{eqn:hyb_ph} given by
\begin{equation}\label{eqn:hyb_ph_real}
-\IIm[g_\lambda^{\text{R}}(\omega)] = \frac{\pi}{2D}\rho_{\text{FD}}^{}(\omega, D, T_{\text{fict}}^{})\rho_{\text{FD}}^{}(-\omega, D, T_{\text{fict}}^{}),
\end{equation}
where $T_{\text{fict}}^{}$ is a fictious temperature used to smoothen the band edge. We choose $T_{\text{fict}}^{} = 0.5 \Gamma$, and the half-bandwidth $D = 10 \Gamma$ to be consistent with Ref.~\cite{do.ga.15} and allow for a comparison of the results. Since the leads are considered in equilibrium, their Keldysh components follow from the fluctuation-dissipation theorem in Eq.~\eqref{eqn:fluc_diss}, where the applied voltage enters within the Fermi-Dirac distribution function as chemical potential $\mu_{\text{R}}=\phi/2=-\mu_{\text{L}}$.

\subsection{\label{sec:imp_model_aux}Auxiliary impurity model}
Our goal is to compute the Green's function of the impurity in presence of interactions. In AMEA one replaces the infinite bath with an open system consisting of a limited number of bath sites coupled to Markovian environments. Its auxiliary hybridization function $\matK{\Delta}_{\text{aux}}^{}$ is determined to resemble the physical one $\matK{\Delta}_{\text{ph}}^{}$ as close as possible, since in the limit $\matK{\Delta}_{\text{aux}}^{} \to \matK{\Delta}_{\text{ph}}^{}$ the solution of the impurity problem becomes exact. For convenience, we choose the impurity to be located at the central site $i=0\eqqcolon f$ connected to two equally long chains with $N_{\text{B}}$ bath sites at $i=\pm 1, \pm 2,\ldots,\pm N_{\text{B}}/2$. The dynamics of the density matrix $\rho$ of this open system is described by the Lindblad equation
\begin{equation}\label{eqn:Lindlad}
\begin{split}
\dot{\rho}(t) &= L\rho \\
&= -\ii[H_{\text{aux}}^{}, \rho] + \sum_{ij\sigma} \Gamma_{ij}^{(1)} \left ( c_{j\sigma}^{} \rho c_{i\sigma}^\dagger - \frac{1}{2} \{c_{i\sigma}^\dagger c_{j\sigma}^{}, \rho \} \right ) \\
&\phantom{= -\ii[H_{\text{aux}}, \rho] }+ \sum_{ij\sigma} \Gamma_{ij}^{(2)} \left ( c_{i\sigma}^\dagger \rho c_{j\sigma}^{} - \frac{1}{2} \{c_{j\sigma}^{} c_{i\sigma}^\dagger, \rho \} \right ),
\end{split}
\end{equation}
where $[A,B]$ and $\{A,B\}$ denote the commutator and anticommutator, respectively. Within the Lindbladian $L$, the unitary time evolution is determined by
\begin{equation} \label{eqn:H_aux}
H_{\text{aux}}^{} = \sum_{ij\sigma} E_{ij}^{} c_{i\sigma}^\dagger c_{j\sigma}^{} + U n_{f\uparrow}^{}n_{f\downarrow}^{}.
\end{equation}
In Fig.~\ref{fig:aux_system}, the \textquoteleft normal space\textquoteright\, part illustrates the action of the Lindbladian. There, the green lines represent the dissipative part of the Lindbladian in form of $\Gamma_{ij}$, while the blue lines denote the unitary part given by the hopping in $H_{\text{aux}}$. The coupling constants $\Gamma_{ij}$, on-site energies $E_{ii}$~\footnote{The on-site energy at the impurity is not a fit parameter, but fixed by particle hole symmetry to $E_{ff}=-U/2$.}, and hoppings $E_{ij}$ occuring in Eq.~\eqref{eqn:Lindlad} are used as parameters to fit the expression for $\matK{\Delta}_{\text{aux}}^{}$ to the data $\matK{\Delta}_{\text{ph}}^{}$. Since the baths in the auxiliary system are noninteracting, one has access to an analytical expression for the auxiliary hybridzation function via the Dyson equation
\begin{equation}
\matK{G}_{\text{0},ff}^{-1} = \matK{g}_{\text{0},ff}^{-1} - \matK{\Delta}_{\text{aux}}
\end{equation}
which in components reads
\begin{equation} \label{eqn:hyb_aux}
\begin{split}
\Delta_{\text{aux}}^\text{R}(\omega) &= 1/g_{\text{0},ff}^\text{R}(\omega) - 1/G_{\text{0},ff}^\text{R}(\omega), \\
\Delta_{\text{aux}}^\text{K}(\omega) &= G_{\text{0},ff}^\text{K}(\omega)/|G_{\text{0},ff}^\text{R}(\omega)|^{2}.
\end{split}
\end{equation}
Here, $G_{\text{0},ff}^\text{R/K}$ are the $ff$ components of the Green's function matrices~\cite{do.so.17}
\begin{equation} \label{eqn:g_0_aux}
\begin{split}
\mat{G}_\text{0}^\text{R}(\omega) &= [\omega - \mat{E} + \ii (\mat{\Gamma}^{\text{(1)}} + \mat{\Gamma}^{\text{(2)}})]^{-1}, \\
\mat{G}_\text{0}^\text{K}(\omega) &= 2 \ii \mat{G}_\text{0}^\text{R}(\omega)(\mat{\Gamma}^{\text{(2)}} - \mat{\Gamma}^{\text{(1)}})\mat{G}_\text{0}^\text{A}(\omega),
\end{split}
\end{equation}
with
\begin{equation}
g_{0,ff}^\text{R}(\omega) = (\omega-\varepsilon_{\text{imp}})^{-1}.\\
\end{equation}
Given the hybridization function of the physical system in Eq.~\eqref{eqn:Delta_ph}, one defines a cost function
\begin{align}
\chi(\mat{E},\mat{\Gamma}^{\text{(1)}},\mat{\Gamma}^{\text{(2)}}) &= \sum_{\mathclap{\alpha\in\{\text{R},\text{K}\}}}\quad\int_{\mathrlap{-\infty}}^{\mathrlap{\infty}}\dd\omega\, W^{\alpha}(\omega)\times \notag \\
&\hspace{-1.0cm}\times \IIm[\Delta_{\text{ph}}^{\alpha}(\omega) - \Delta_{\text{aux}}^{\alpha}(\omega;\mat{E},\mat{\Gamma}^{\text{(1)}},\mat{\Gamma}^{\text{(2)}})]^{2}
\end{align}
to assess the quality of the fit with the hybridization function of the auxiliary system. In this paper, the weight function $W^{\alpha}(\omega)=\Theta(|\omega|-\omega_{\text{max}})$ only restricts the frequency range with $\omega_{\text{max}}=15 \Gamma$. More aspects of the fitting procedure are discussed in Ref.~\cite{do.so.17}. The aforementioned paper also shows that in case all available fit parameters are used, the shape of the bath in the auxiliary system does not affect the quality of the fit. 
This justifies our choice of bath geometry.

\subsection{\label{sec:superferm}Superfermion representation}
In the following, we sketch how the superfermion representation~\cite{dz.ko.11} serves as a convenient way to express the Lindblad equation. For an alternative, equivalent representation, see~\cite{pros.08}. Within this section, we roughly follow the presentations in~\cite{do.nu.14,ar.do.18}. 

The basic idea is to vectorize the density matrix and describe it in terms of a state that lives in a doubled Hilbert space to convert the Lindblad equation into a linear algebra problem. The usual density matrix can be written as
\begin{equation} \label{eqn:density_matrix_expanded}
\rho = \sum_{mn} \rho_{mn}\ket{m}\bra{n}.
\end{equation}
\begin{figure}[t]
\centering
\includegraphics[width=\linewidth]{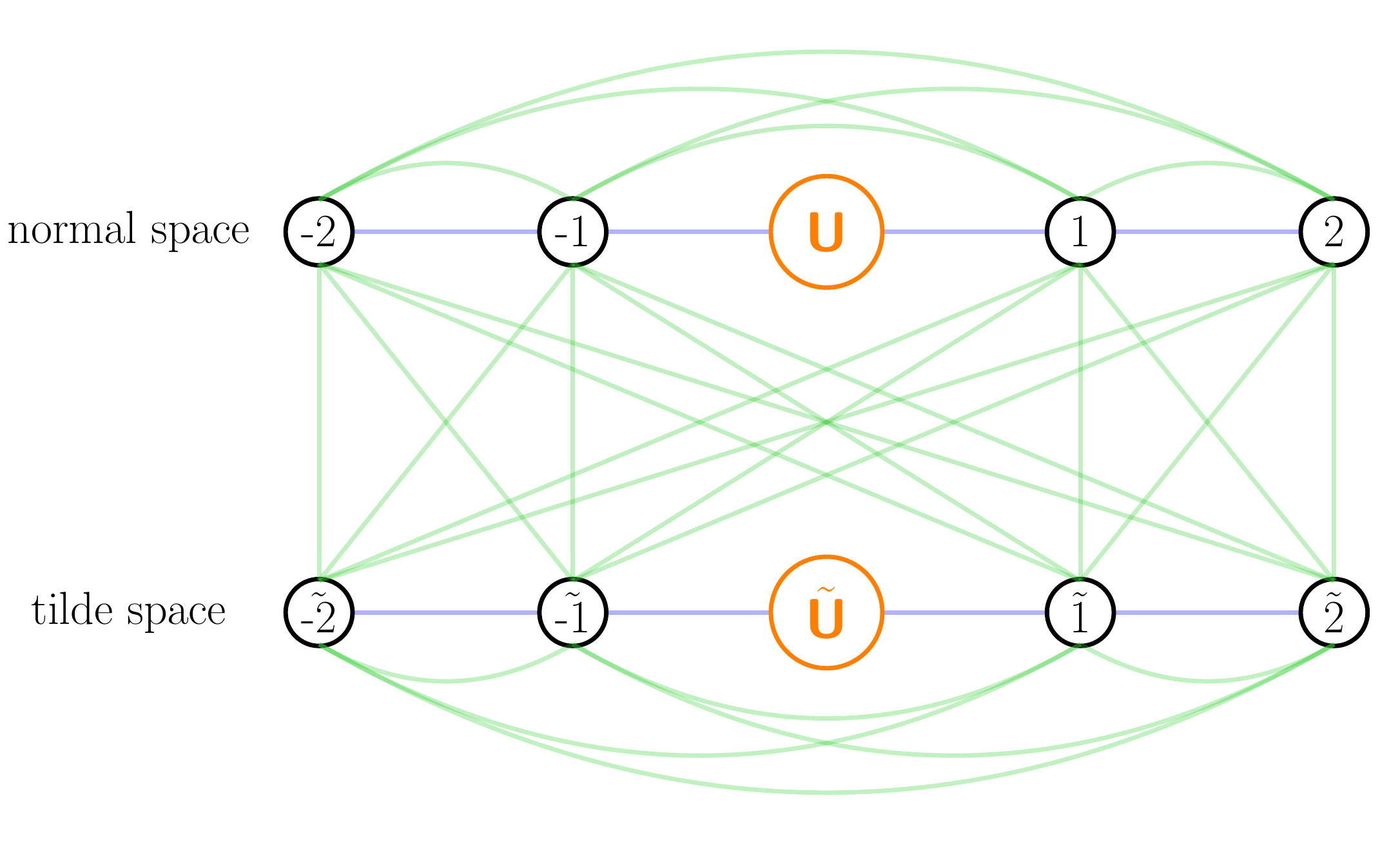}
\caption{Auxiliary system in the superfermion representation. The green lines represent the dissipative part $\mat{\Gamma}^{(1/2)}$ of the Lindbladian and the blue ones the unitary part $\mat{E}$.}
\label{fig:aux_system}
\end{figure}
To transform the bra state into a ket state, one introduces the left vacuum~\footnote{Note that the state $\ket{m}\otimes \ket{\tilde{m}}$ contains a phase factor~\cite{ar.do.18}.}
\begin{equation} \label{eqn:left_vacuum}
\ket{I} = \sum_m \ket{m} \otimes \ket{\tilde{m}}.
\end{equation}
With this, one can transform the density matrix into a state vector as
\begin{equation} \label{eqn:density_matrix_superfermion}
\begin{split}
\rho \to \ket{\rho} &= (\rho \otimes \tilde{\mathds{1}}) \ket{I} \\
&= \left ( \sum_{mn} \rho_{mn} \ket{m}\bra{n} \otimes \tilde{\mathds{1}} \right ) \left( \sum_{j} \ket{j} \otimes \ket{\tilde{j}} \right) \\
&= \sum_{mnj} \rho_{mn} (\ket{m}\braket{n|j}) \otimes \ket{\tilde{j}} = \sum_{mn} \rho_{mn} \ket{m}\otimes\ket{\tilde{n}}.
\end{split}
\end{equation}
By applying the Lindblad equation to the left vacuum, one can express it in the superfermion representation. The ensuing aim is to move the density matrix to the left vacuum, such that one can employ Eq.~\eqref{eqn:density_matrix_superfermion} to turn it into a state. This requires commutations of the density matrix with creation and annihilation operators. To achieve this, one uses that the left vacuum transfers the creation (annihilation) operators to the tilde space via
\begin{equation} \label{eqn:tilde_conj_rel}
\begin{split}
c_j^\dagger \ket{I} &= - \ii \tilde{c}_j^{} \ket{I}, \\
c_j \ket{I}^{} &= - \ii \tilde{c}_j^\dagger \ket{I},
\end{split}
\end{equation}
which then commute with the density matrix~\footnote{Fermionic operators from the normal- and tilde space anticommute~\cite{dz.ko.11}. Since the density matrix contains even products of normal-space fermionic operators~\cite{ar.do.18}, it commutes with the operators in tilde-space.}. The Lindbladian then follows as
\begin{equation} \label{eqn:Lindblad_superferm}
\begin{split}
\ii L &= \sum_\sigma \vec{c}_\sigma^\dagger \mat{h} \vec{c}_\sigma^{} -2 \Tr(\mat{E}+\ii\mat{\Lambda}) \\
&+U \left ( n_{f\uparrow}^{} n_{f\downarrow}^{} - \tilde{n}_{f\uparrow}^{} \tilde{n}_{f\downarrow}^{} \right ),
\end{split}
\end{equation}
with the matrix
\begin{equation} \label{eqn:h_matrix}
\begin{split}
\mat{h} = \left(\begin{matrix}
\mat{E} + \ii \mat{\Omega} & \phantom{-}2 \mat{\Gamma}^{\text{(2)}} \\
-2 \mat{\Gamma}^{\text{(1)}} & \mat{E} - \ii \mat{\Omega}
\end{matrix}\right),
\end{split}
\end{equation}
the vector $\vec{c}_\sigma^\dagger = (c_{0\sigma}^\dagger,\ldots , c_{N_{\text{B}}\sigma}^\dagger, \tilde{c}_{0\sigma}^{},\ldots , \tilde{c}_{N_{\text{B}}\sigma}^{})$ as well as the matrices $\mat{\Omega} = \mat{\Gamma}^{\text{(2)}} - \mat{\Gamma}^{\text{(1)}}$ and $\mat{\Lambda} = \mat{\Gamma}^{\text{(2)}} + \mat{\Gamma}^{\text{(1)}}$. Note that $f=0$ refers to the impurity site. The action of the matrices $\mat{E}$, $\mat{\Gamma}^{(1)}$ and $\mat{\Gamma}^{(2)}$ in the superfermion representation are illustrated in Fig.~\ref{fig:aux_system}. It can be seen, that the spin-resolved difference in particle number between normal- and tilde-space
\begin{equation} \label{eqn:conservation}
\hat{N}_{\sigma} - \hat{\tilde{N}}_{\sigma} = \sum_{i} (c_{i\sigma}^{\dagger}c_{i\sigma}-\tilde{c}_{i\sigma}^{\dagger}\tilde{c}_{i\sigma})
\end{equation}
is conserved~\cite{ar.kn.13}. Since the left vacuum is the left eigenstate corresponding to the steady state, and it lies within the subspace with $N_{\sigma}-\tilde{N}_{\sigma} = 0$~\footnote{The left vacuum given in Eq.~(\ref{eqn:left_vacuum}) is the sum over all combinations of normal and tilde space states with the same particle number, i.e. $N=\tilde{N}$.}, also the steady state must lie in this subspace.

The interacting Green's function can be obtained in its Lehmann representation. The Fourier transform of the greater and lesser component ($\gtrless$) obtained for positive times ($+$) read
\begin{equation} \label{eqn:Lehmann_greater}
\begin{split}
G_{ij}^{>+}(\omega) &= \sum_k \bra{I} c_i^{} \ket{k\text{R}} \bra{k\text{L}} c_j^\dagger \ket{\rho_\infty} \frac{1}{\omega - \ii L_k^{}}, \\
G_{ij}^{<+}(\omega) &= \sum_k \bra{I} c_j^{\dagger} \ket{k\text{R}} \bra{k\text{L}} c_i^{} \ket{\rho_\infty} \frac{1}{\omega + \ii L_k^{}}
\end{split}
\end{equation}
with the left and right eigenvectors $\ket{k\text{L}}$ and $\ket{k\text{R}}$ as well as eigenvalues $L_{k}$ of the Lindbladian $L$~\footnote{The notation used here is taken partially from Ref.~\cite{do.nu.14}, except for the factor $\ii=\sqrt{-1}$ in the denominator. This is because $\ket{k\text{L}}$, $\ket{k\text{R}}$ and $L_{k}$ are the eigenvectors and eigenvalues of the Lindbladian $L$.}. Using the relations
\begin{equation} \label{eqn:GF_relations}
\begin{split}
G_{ij}^{\gtrless -}(\omega) &= -[G_{ji}^{\gtrless +}(\omega)]^{\ast}, \\
G_{ij}^{\text{R}}(\omega) &= G_{ij}^{>+} - G_{ij}^{<-}, \\
G_{ij}^{\text{K}}(\omega) &= G_{ij}^{>+} + G_{ij}^{<+} - G_{ij}^{>-} - G_{ij}^{<-}, \\
\Sigma^{\text{R}}(\omega) &= 1/G_\text{0}^\text{R}(\omega) - 1/G_{}^\text{R}(\omega), \\
\Sigma^{\text{K}}(\omega) &= -G_\text{0}^K(\omega)/|G_\text{0}^\text{R}(\omega)|^2 + G_{}^\text{K}(\omega)/|G_{}^\text{R}(\omega)|^2,
\end{split}
\end{equation}
one obtains the full Green's function and selfenergy which can be used to compute other physical observables.

In practice, we express the Lindbladian $L$ in a given basis and obtain the steady state $\ket{\rho_{\infty}}$ using the biconjugate gradient method~\cite{meis}. We construct the basis using CI as described in Sec.~\ref{sec:CI}. After applying the creation/annihilation operator to the steady state, we compute the matrix elements contributing to the Green's function and the eigenvalues $L_{k}$ of the Lindbladian using the Bi-Lanczos method~\cite{meis}. With the assembled quantities defined in Eq.~\eqref{eqn:Lehmann_greater}, we obtain the Green's function and selfenergy via Eq.~\eqref{eqn:GF_relations}.

\subsection{\label{sec:CI}Configuration interaction expansion}
The configuration interaction (CI) expansion is well known for the equilibrium case~\cite{he.jo, sh.sc.99, zg.gu.12, li.de.13} and has been used for nonequilibrium situations as well~\cite{dz.ko.14}. 

In equilibrium, one first determines the many-body ground state for an effectively noninteracting system, e.g. within a mean field approximation. It serves as a reference state to construct a restricted basis in which the Hamiltonian is diagonalized. This basis of many-body states is obtained by applying particle-hole (PH) excitations to the reference state, meaning the removal of a particle from one single-particle state and addition of a particle to another one.

Out of equilibrium, the many-body steady state for an effectively noninteracting system is used as reference state and the Lindbladian replaces the Hamiltonian. Within the superfermion representation (see Sec.~\ref{sec:superferm}), the basis is generated by changing the occupation of the single-particle states labeled by the normal modes~\cite{do.nu.14} of the Lindbladian of the effectively noninteracting system, which will be introduced below. We will call the excitations used to generate the basis states also PH excitations~\footnote{Note that the PH excitation introduced in Eq.~(\ref{eqn:ph_excitation_xi}) preserves the difference $N_{\sigma}-\tilde{N}_{\sigma}$ rather than the total \textquoteleft particle\textquoteright\, number $N_{\sigma}+\tilde{N}_{\sigma}$. However motivated by the tilde conjugation rules in Eq.~(\ref{eqn:tilde_conj_rel}), particles in the tilde-space can be considered as holes in normal-space and vice versa. From this perspective, the PH excitation $\Bar{\xi}_{i}\xi_{j}\sim c_{k}^{\dagger}\tilde{c}_{l}^{\dagger}, \tilde{c}_{l}^{}c_{k}^{}, c_{k}^{\dagger}c_{k}^{}, \tilde{c}_{l}^{}\tilde{c}_{l}^{\dagger}$ operating in normal-space creates a particle and a hole, annihilates a particle and a hole, creates a particle and annihilates another one, and lastly creates a hole and annihilates another one. It thus preserves the actual particle number before going over to the superfermion representation.}. For the remainder of this paper, single-particle states are denoted as \textquoteleft orbitals\textquoteright\, and many-body states are shortened to \textquoteleft states\textquoteright.

The noninteracting system is constructed by decoupling the interaction term similarly to the Hartree-Fock (HF) approximation via
\begin{equation} \label{eqn:hartree_fock}
\begin{split}
n_{f\uparrow}n_{f\downarrow} \to \text{HF}(\braket{n_{f\uparrow}})&=n_{f\uparrow}\braket{n_{f\downarrow}}+\braket{n_{f\uparrow}}n_{f\downarrow}\\
&-\braket{n_{f\uparrow}}\braket{n_{f\downarrow}}, \\
\tilde{n}_{f\uparrow}\tilde{n}_{f\downarrow} \to \widetilde{\text{HF}}(\braket{\tilde{n}_{f\uparrow}})&=\tilde{n}_{f\uparrow}\braket{\tilde{n}_{f\downarrow}}+\braket{\tilde{n}_{f\uparrow}}\tilde{n}_{f\downarrow}\\
&-\braket{\tilde{n}_{f\uparrow}}\braket{\tilde{n}_{f\downarrow}}.
\end{split}
\end{equation}
As we will show in Sec.~\ref{sec:dof}, in practice it is more suitable to treat the expectation values $\braket{n_{f\uparrow}}$, $\braket{n_{f\downarrow}}$, $\braket{\tilde{n}_{f\uparrow}}$, $\braket{\tilde{n}_{f\downarrow}}$ as free parameters and fix them via an explicit parameter sweep rather than to determine them self-consistently. Due to the conservation of Eq.~\eqref{eqn:conservation} and since we work (on average) at half filling, we can restrict ourselves to
\begin{equation} \label{eqn:parameter_space}
\begin{split}
\braket{\tilde{n}_{f\sigma}} &= \braket{n_{f\sigma}}, \\
\braket{n_{f\downarrow}} &= 1-\braket{n_{f\uparrow}}, \\
m_{\uparrow}\coloneqq\braket{n_{f\uparrow}} &\in[0,0.5].
\end{split}
\end{equation}
The constant offset in the HF approximation cancels in the Lindbladian of Eq.~\eqref{eqn:Lindblad_superferm}. When absorbing the HF contribution, the matrix $\mat{h}$ introduced in Eq.~\eqref{eqn:h_matrix} picks up a spin-dependence $\mat{h}_{\sigma}$ in the form of
\begin{equation}
\begin{split}
E_{ii\sigma} = E_{ii} + U\braket{n_{f\Bar{\sigma}}}\delta_{if},
\end{split}
\end{equation}
where $\Bar{\sigma}$ denotes the spin-direction opposite to $\sigma$ and the second term in Eq.~(\ref{eqn:Lindblad_superferm}) becomes
\begin{equation}
\eta=-\sum_{\sigma}^{} \Tr(\mat{E}_{\sigma})-2\ii\Tr(\mat{\Lambda}).
\end{equation}
The non-Hermitian matrix $\mat{h}_{\sigma}$ can be diagonalized~\cite{do.nu.14} as
\begin{equation} \label{eqn:diagonal_op_mat}
\mat{\varepsilon} = \mat{V}^{-1} \mat{h}_{\sigma} \mat{V}
\end{equation}
with a diagonal matrix $\mat{\varepsilon}$, so that the effective noninteracting Lindbladian amounts to
\begin{equation} \label{eqn:Lindblad_non_int}
\ii L_0 = \sum_{\sigma}\Bar{\vec{\xi}} \mat{\varepsilon} \vec{\xi} + \eta.
\end{equation}
Here and in the following the spin-index $\sigma$ is suppressed. The operators $\Bar{\vec{\xi}}=\vec{c}^\dagger \mat{V}$ and $\vec{\xi}=\mat{V}^{-1} \vec{c}$ denote the \textquoteleft creation\textquoteright\, and \textquoteleft annihilation\textquoteright\, operators labeled by the normal modes which obey the canonical anticommutation relations
\begin{equation} \label{eqn:anticommutation}
\{\xi_{i},\Bar{\xi}_{j}\} = \delta_{ij},
\end{equation}
but are not mutually Hermitian conjugate. 

To motivate how these orbitals labeled by normal modes allow for the notion of a PH excitation and which quantity corresponds to the single-particle energies, we follow Ref.~\cite{do.nu.14}. The steady state is time-independent
\begin{equation} \label{eqn:steady_state_nonint}
L_{0} \ket{\rho_{\infty 0}} = 0.
\end{equation}
In contrast, the time evolution of a normal mode operator applied to the steady state depends on time
\begin{equation} \label{eqn:time_evolution}
\ee^{L_0 t} \xi_{i} \ket{\rho_{\infty 0}} = \ee^{L_0 t} \xi_{i} \ee^{-L_0 t} \ket{\rho_{\infty 0}} = \ee^{\ii\varepsilon_{i} t} \xi_{i} \ket{\rho_{\infty 0}}.
\end{equation}
Since $\IIm(\varepsilon_{i})<0$ implies a diverging state, while the long-time limit is given by the steady state itself, we must have
\begin{equation}
\xi_{i} \ket{\rho_{\infty 0}} = 0\quad\text{ for }\IIm(\varepsilon_{i})<0.
\end{equation}
Similarly it must hold that
\begin{equation}
\Bar{\xi}_{i} \ket{\rho_{\infty 0}} = 0\quad\text{ for }\IIm(\varepsilon_{i})>0.
\end{equation}
The steady state can thus be interpreted as a kind of \textquoteleft Fermi sea\textquoteright, where $\IIm(\varepsilon_{i})$ takes the role of the single-particle energies separating the \textquoteleft filled\textquoteright\, orbitals at $\IIm(\varepsilon_{i})>0$ from the \textquoteleft empty\textquoteright\, orbitals at $\IIm(\varepsilon_{i})<0$. Within this line of argument, a PH excitation is given by
\begin{equation} \label{eqn:ph_excitation_xi}
\Bar{\xi}_{i}\xi_{j} \ket{\rho_{\infty 0}}.
\end{equation}
with the indices $i,j$ chosen such that the state is nonvanishing.
\begin{figure}[t]
\centering
\includegraphics[width=\linewidth]{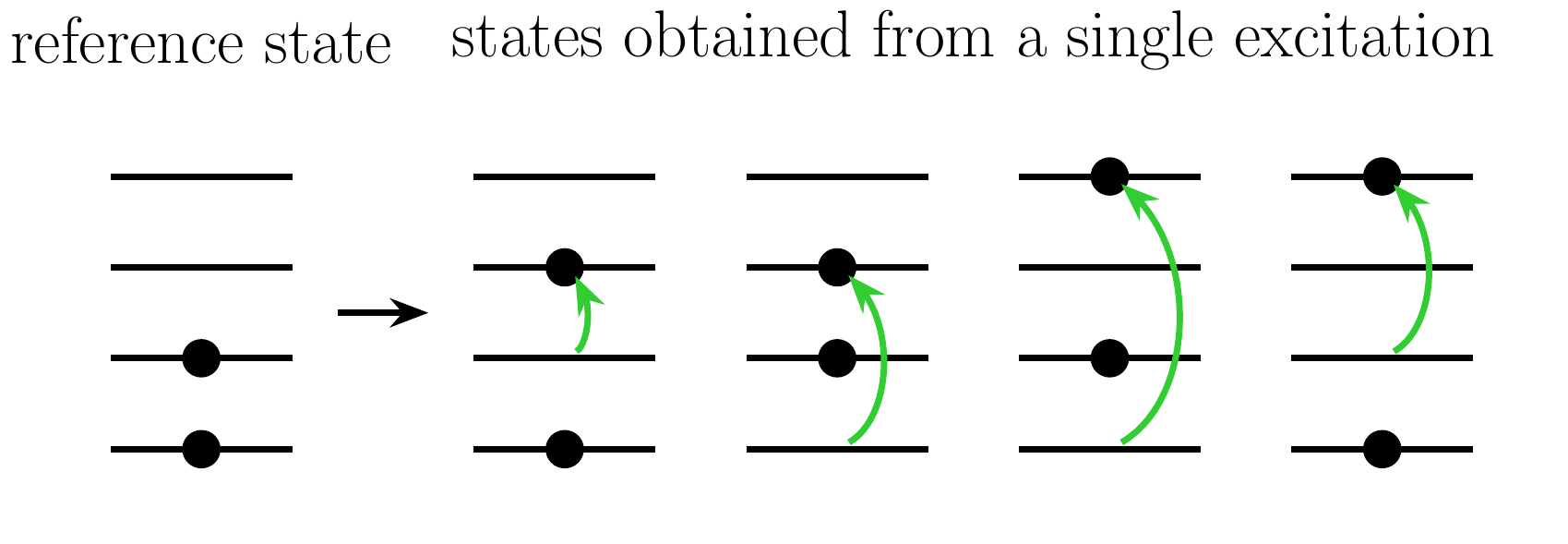}
\caption{Construction of the many-body basis in CI using single PH excitations for the orbitals labeled by normal modes. We start from the reference state and add all states attainable with a single PH excitation. The green arrows highlight the annihilation of a particle in an occupied orbital and creation of a particle in an empty orbital. Here the construction omits the state highlighted in green in Fig.~\ref{fig:CAS} which is part of the full basis. With double excitations, we would reach the full state space.}
\label{fig:CI}
\end{figure}

Basis states are expressed in terms of the occupations $\Bar{\xi}_{i}\xi_{i}$ of the orbitals labeled by normal modes. As discussed above, the reference state is a \textquoteleft Fermi sea\textquoteright, where $\IIm(\varepsilon_{i})$ separates filled from empty orbitals. Applying PH excitations given in Eq.~\eqref{eqn:ph_excitation_xi} to the reference state generates further basis states~\footnote{The states generated in CI resemble Slater determinants due to the canonical anticommutation relations in Eq.~(\ref{eqn:anticommutation}).} as illustrated in Fig.~\ref{fig:CI}. The maximum number of allowed PH excitations on top of the reference state determines the dimension of the basis and thereby the restriction of the Hilbert space. Since each excitation increases the size of the Hilbert space exponentially, we will restrict ourselves to three PH excitations~\footnote{Here CI refers to CISDT in terms of the conventional language~\cite{he.jo, sh.sc.99}.}. 

An extension of CI consists of using more reference states by means of a so-called complete active space (CAS). Results of such an extension (CASCI), which is computationally more costly, turn out not to improve significantly  our CI results. For this reason, we present its implementation and results in App.~\ref{sec:CASCI}.

It is convenient to further simplify the noninteracting Lindbladian by performing a PH transformation. It serves to make the steady state become the vacuum and thereby eliminate the constant term $\eta$. The transformation reads
\begin{equation} \label{eqn:op_transformation}
\begin{split}
\vec{P} &= \Bar{\mat{D}} \vec{\xi} + \Bar{\vec{\xi}} \mat{D}, \\
\Bar{\vec{P}} &= \mat{D} \vec{\xi} + \Bar{\vec{\xi}} \Bar{\mat{D}}
\end{split}
\end{equation}
with components
\begin{equation} \label{eqn:D_imag}
\begin{split}
D_{ij}^{} &= \delta_{ij}^{} \Theta[\IIm(\varepsilon_i^{})], \\
\Bar{D}_{ij}^{} &= 1-D_{ij}^{}.
\end{split}
\end{equation}
Considering the creation operators $\Bar{\xi}_{i}$, particles stay particles if $\IIm(\varepsilon_{i})<0$ and become holes when $\IIm(\varepsilon_{i})>0$. After the transformation, the Lindbladian becomes
\begin{equation} \label{eqn:Lindblad_p_basis}
\ii L_{0}^{} = \sum_i \varepsilon_i^{} \Bar{P}_i^{} P_i^{}.
\end{equation}
Note that when $\xi$-creation operators become $P$-annihilation operators and vice versa, we obtain a factor $-1$ for commutating, which we absorb in $\varepsilon$. Since the constant $\eta$ is removed, every $P_i^{}$ annihilates the steady state and every $\Bar{P}_i^{}$ annihilates the left vacuum. After the transformation, a PH excitation reads
\begin{equation} \label{eqn:ph_excitation_p}
\Bar{P}_{i}\Bar{P}_{j} \ket{\rho_{\infty 0}}.
\end{equation} 

The interacting Lindbladian finally follows as
\begin{equation} \label{eqn:Lindblad_int}
\begin{split}
\ii L &= \sum_i \varepsilon_i^{} \Bar{P}_i^{} P_i^{} - [\text{HF}(\braket{n_{f\uparrow}^{}})-\widetilde{\text{HF}}(\braket{\tilde{n}_{f\uparrow}^{}})]\\
&+ U \left( n_{f\uparrow}^{} n_{f\downarrow}^{} - \tilde{n}_{f\uparrow}^{} \tilde{n}_{f\downarrow}^{} \right).
\end{split}
\end{equation}
To take the correlation part properly into account, the decoupled interaction term is removed from the effective noninteracting Lindbladian. Expressing the correlation part in terms of $P$ and $\Bar{P}$ leads to additional terms. The choice of the free parameter $m_{\uparrow}\coloneqq\braket{n_{f\uparrow}}$ contained in the effective noninteracting Lindbladian will be discussed in Sec.~\ref{sec:dof}.

\begin{table*}[t]
\caption{\label{tab:dimension_hilbert_space}Dimension of the Hilbert spaces $\dim(\mathcal{H})$ encountered when computing the steady state $\ket{\rho_{\infty}}$ and the Green's function $G$ for different numbers of bath sites $N_{\text{B}}$ employing CI and CASCI compared to the full Hilbert space. The conservation of $N_{\sigma}-\tilde{N}_{\sigma}=0, \pm 1$ for $\ket{\rho_{\infty}}$ and $G$ respectively is exploited for all cases.}
\begin{ruledtabular}
\begin{tabular}{lrrrrrr}
$N_{\text{B}}$ & $\dim(\mathcal{H}_{\ket{\rho_{\infty}}})$ CI & $\dim(\mathcal{H}_{\ket{\rho_{\infty}}})$ CASCI & $\dim(\mathcal{H}_{\ket{\rho_{\infty}}})$ full & $\dim(\mathcal{H}_{G})$ CI & $\dim(\mathcal{H}_{G})$ CASCI & $\dim(\mathcal{H}_{G})$ full\\
2  & 282     & 356       & 400                & 522       & 584       & 600 \\
4  & 6 076   & 10 980    & 63 504             & 20 160    & 32 924    & 105 840 \\
6  & 49 050  & 104 356   & 11 778 624         & 244 328   & 485 748   & 20 612 592 \\
8  & 233 380 & 540 996   & 2 363 904 400      & 1 561 860 & 3 445 588 & 4 255 027 920 \\
10 & 807 434 & 1 971 684 & 497 634 306 624    & 6 799 252 & --        & 912 329 562 144 \\
\end{tabular}
\end{ruledtabular}
\end{table*}

As established before, the noninteracting steady state expressed in $P$- and $\Bar{P}$-operators is the vacuum state. To construct the basis within CI for the computation of the interacting steady state, we apply two, four and six $\Bar{P}$-operators to the noninteracting steady state, which correspond to one, two and three PH excitations. After filtering the resulting states by requiring them to lie in the subspace with $N_{\sigma}-\tilde{N}_{\sigma} = 0$, we use the basis to construct the entries of the Lindbladian in Eq.~\eqref{eqn:Lindblad_int} as mentioned in Sec.~\ref{sec:superferm}. For the Green's function in CI, we generate the basis of excited states by applying one, three, five and seven $\Bar{P}$-operators to the noninteracting steady state as required by Eq.~\eqref{eqn:Lehmann_greater}. Here, we filter the states by requiring them to lie in the subspace with $N_{\sigma}-\tilde{N}_{\sigma} = \pm 1$. Since the number of these states is quite large, as seen in Tab.~\ref{tab:dimension_hilbert_space}, we restrict the basis even further. More specifically, in the expansion
\begin{equation}
c_{i}^{\dagger}\ket{\rho_{\infty}} = \sum_k \beta_{k}\ket{x_{k}}
\end{equation}
in terms of the basis states $\ket{x_{k}}$ generated using one, three, five and seven $\Bar{P}$-operators, we remove $75\%$ of the states with the smallest coefficients $|\beta_{k}|$.

\section{\label{sec:results}Results}
\subsection{\label{sec:dof}Method parameter}
Let us begin by discussing the free parameter of CI. To fully determine our interacting Lindbladian given in Eq.~\eqref{eqn:Lindblad_int}, we have to fix the parameter $m_{\uparrow}\coloneqq\braket{n_{f\uparrow}}$ whose range is specified in Eq.~\eqref{eqn:parameter_space}. In practice it turns out that whenever the fit matches the hybridization function well, this parameter has little effect on the results. It becomes relevant for more challenging cases with sharp features in the hybridization function, e.g. at low temperatures and for strong correlations within DMFT, where correlations have a backaction onto the hybridization function.

Since we are most interested in the regime of low temperature and high electron-electron interaction, within this section we set $T/\Gamma=0.05$ and $U/\Gamma=6$. Our benchmark to adjust the parameters of the method are ED calculations, which are based on the same set of fit parameters. Hence any deviation between CI and ED emerges from the approximative character of CI. We use $N_{\text{B}}=6$ bath sites, since this is the upper limit for ED.

\begin{figure}[b]
\centering
\includegraphics[width=0.9\linewidth]{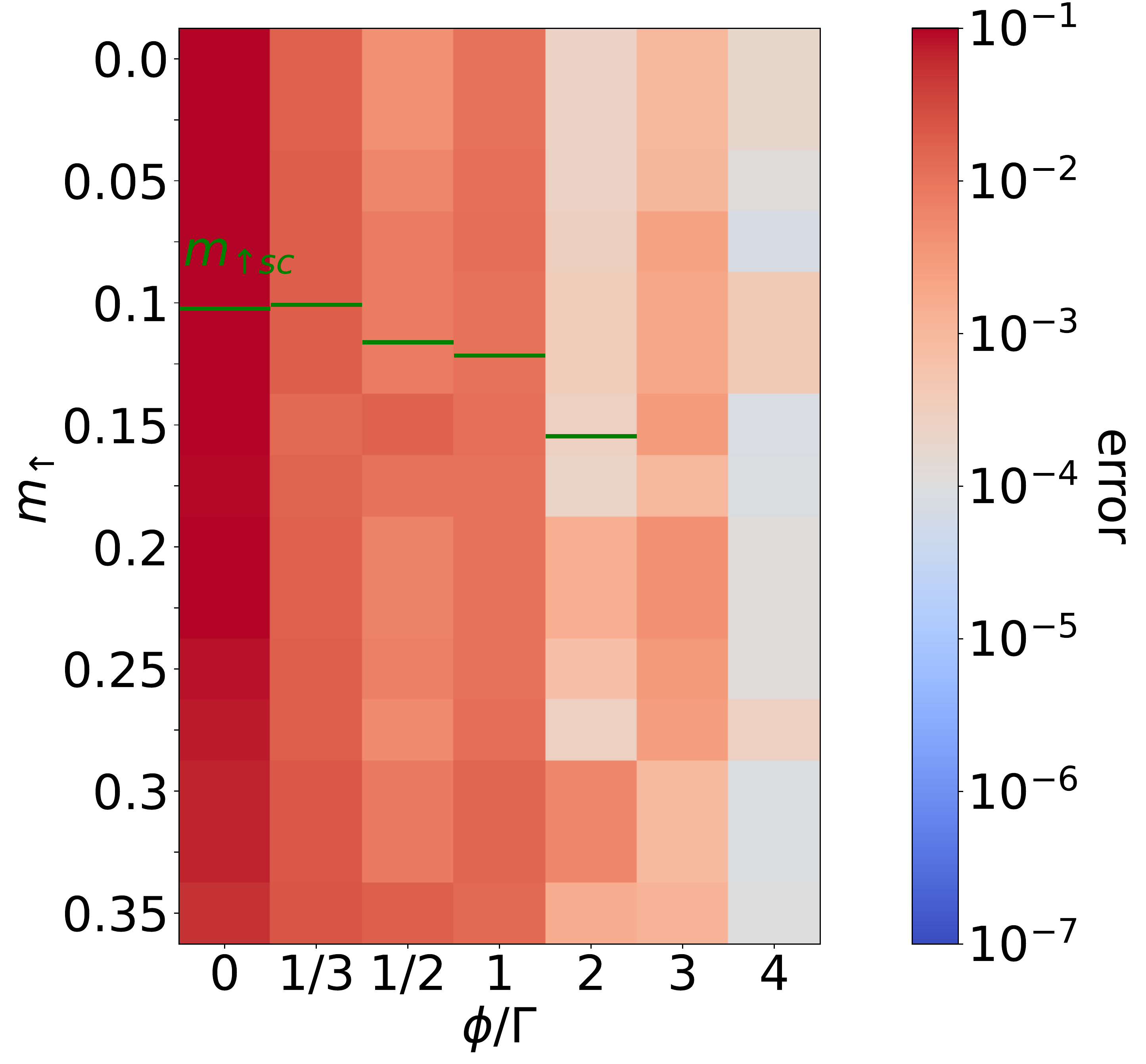}
\caption{Matrix plot of the error defined in Eq.~\eqref{eqn:error_calc} for CI as solver for AMEA given different voltages $\phi$ and parameters $m_{\uparrow}$ with resolution $\Delta m_{\uparrow}=0.025$. For $m_{\uparrow}\in[0.375,0.5]$, the error jumps two orders of magnitude for moderate and large voltages and is omitted here. Green lines denote self-consistently determined parameters $m_{\uparrow\text{sc}}$~\cite{Note9}. The remaining parameters are the temperature $T/\Gamma=0.05$, Hubbard interaction $U/\Gamma=6$, and $N_{\text{B}}=6$ bath sites.}
\label{fig:CI_error_colorplot}
\end{figure}

Figure~\ref{fig:CI_error_colorplot} shows the difference between CI and ED, calculated as
\begin{equation} \label{eqn:error_calc}
\begin{split}
\text{error} = \int_{\mathrlap{-\infty}}^{\mathrlap{\infty}} \dd\omega \bigg{(} &\left\{ \frac{\IIm[G^{\text{K}}_{\text{CI}}(\omega)]_{\phantom{\text{max}}}}{\IIm[G^{\text{K}}_{\text{ED}}(\omega)]_\text{max}} - \frac{\IIm[G^{\text{K}}_{\text{ED}}(\omega)]_{\phantom{\text{max}}}}{\IIm[G^{\text{K}}_{\text{ED}}(\omega)]_\text{max}} \right\}^2 \\
+ &\left\{\frac{\IIm[G^{\text{R}}_{\text{CI}}(\omega)]_{\phantom{\text{max}}}}{\IIm[G^{\text{R}}_{\text{ED}}(\omega)]_\text{max}} - \frac{\IIm[G^{\text{R}}_{\text{ED}}(\omega)]_{\phantom{\text{max}}}}{\IIm[G^{\text{R}}_{\text{ED}}(\omega)]_\text{max}} \right\}^2 \bigg{)},
\end{split}
\end{equation}
where $\IIm[G^{\alpha}_{\beta}(\omega)]_{\text{max}}=\max_{\omega}\{\IIm[G^{\alpha}_{\beta}(\omega)]\}$.
Note that the worst and partially non-convergent results obtained for $m_{\uparrow}>0.35$ around half-filling $m_{\uparrow}=0.5$ are omitted. The HF values $m_{\uparrow\text{sc}}$~\footnote{These are the expectation values in the self-consistently determined HF steady state.}.
perform better than $m_{\uparrow}=0.5$ for small to intermediate voltages, but the minimal difference between ED and CI is located around $m_{\uparrow} = 0.3$~\footnote{Since there occurs a jump in the error by two orders of magnitude starting at $m_{\uparrow} = 0.375$, the small error at $m_{\uparrow} = 0.35$ is probably pathological.}. Therefore, we fix $m_{\uparrow}=0.3$ for the rest of the paper~\footnote{Around this value results do not depend much on the exact choice of $m_{\uparrow}$.}. 
%
Notice that a larger voltage provides a better agreement, so that the equilibrium case seeems the most challenging one. Therefore, we will investigate it in more detail in Sec.~\ref{sec:real_results}. The matrix plot showing the deviations for CASCI can be found in Fig.~\ref{fig:CASCI_error_colorplot} within App.~\ref{subsec:CASCI_results}.

\subsection{\label{sec:real_results}Comparison with NRG and MPS}
In the previous section, we confined ourselves to $N_{\text{B}}=6$ bath sites for CI to use ED as a benchmark. This does not exploit the main advantage of CI to access larger $N_{\text{B}}$ which increases the number of parameters to fit the hybridization function for a more accurate description of the physical impurity model. However there is an opposing effect since the fraction of Hilbert space addressed by CI compared to the full Hilbert space shrinks exponentially fast with increasing $N_{\text{B}}$ as can be inferred from Tab.~\ref{tab:dimension_hilbert_space}. In principle, we can address systems with up to $N_{\text{B}}=10$ bath sites with CI, which provide access to smaller temperatures and slightly more accurate results than with $N_{\text{B}}=8$. However this comes at the expense of a much longer wall time of about $\unit[5]{h}$~\footnote{For our computations, we used 16 threads on an Intel Xeon E5-2650 processor.}. For this reason, we use at most $N_{\text{B}}=8$ bath sites throughout this section. 
\begin{figure}[t]
\centering
\includegraphics[width=\linewidth]{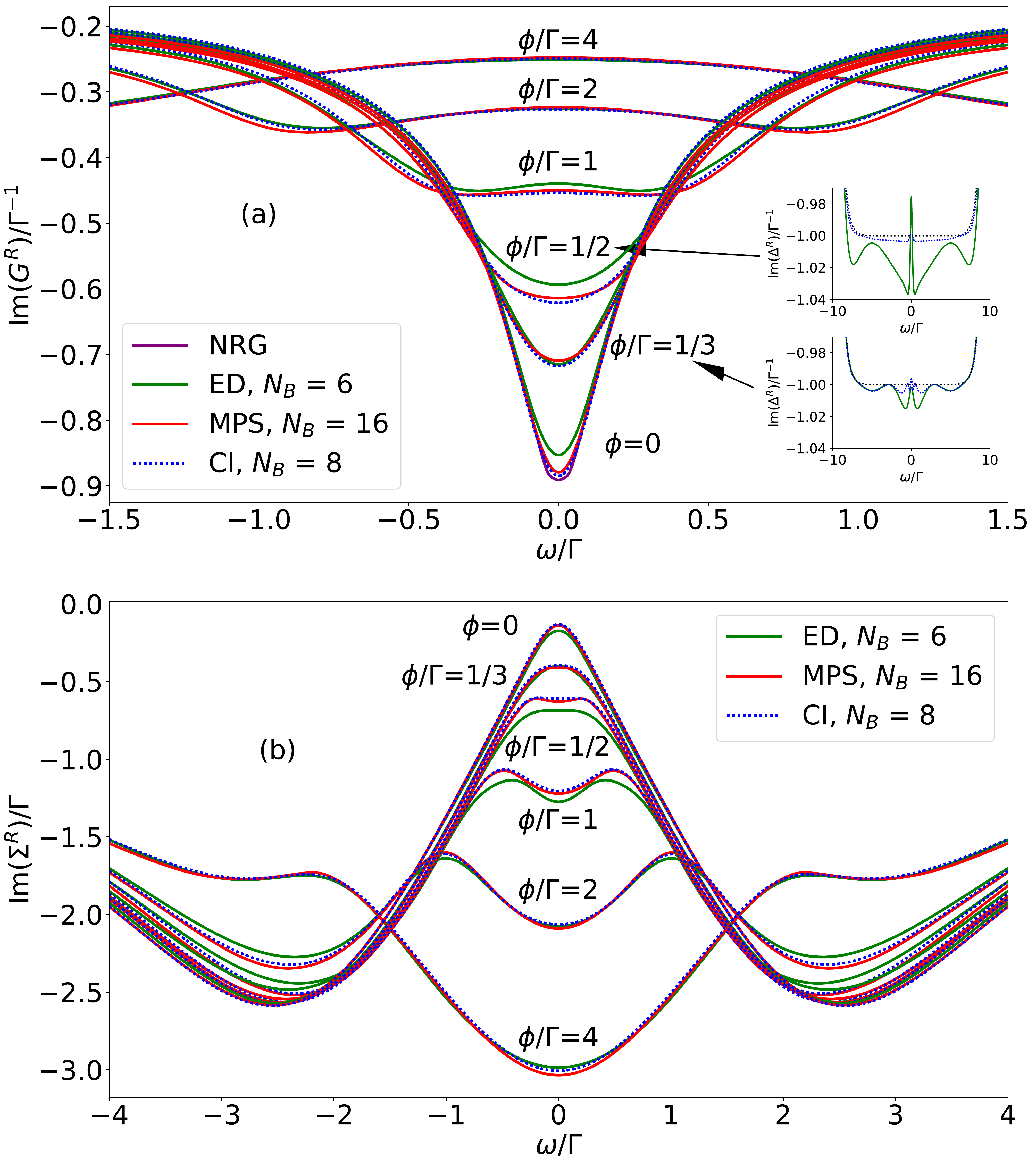}
\caption{Comparison of the imaginary parts of the retarded components of the (a)~Green's function $G$ and the (b)~selfenergy $\Sigma$ for temperature $T/\Gamma=0.05 $, Hubbard interaction $U/\Gamma=6$, $N_{\text{B}}$ bath sites specified in the legend and various voltages $\phi$ obtained from CI, ED and MPS~\cite{do.ga.15} as solvers for AMEA. NRG~\cite{zitk.21} serves as reference for the equilibrium case. The legend of the main figure also applies to the insets, which show the auxiliary hybridization functions for $\phi=1/3, 1/2$ and compares them to the physical hybridization function (black dotted line) as given in Eq.~\eqref{eqn:hyb_ph_real}.}
\label{fig:GF_MPS_CI_compare}
\end{figure}

Figure~\ref{fig:GF_MPS_CI_compare} shows a comparison of $\IIm(G^\text{R})$ and $\IIm(\Sigma^\text{R})$ for $T/\Gamma=0.05$, $U/\Gamma=6$, $N_{\text{B}}$ specified in the legend and various voltages $\phi$ which are computed using CI, ED and MPS as solvers for AMEA. The MPS results are taken from Ref.~\cite{do.ga.15} and serve as our benchmark, since they are the most accurate among the three~\footnote{ED and CI use all available fitting parameters. With $N_{\text{B}}=6$ and $8$, they provide $27$ and $44$ parameters~\cite{do.so.17}. Due to the growth of entanglement, MPS uses only nearest-neighbor couplings. With $N_{\text{B}}=16$ bath sites, it has $46$ fitting parameters available.}. In equilibrium, NRG as implemented in~\cite{zitk.21} serves as reference. For the largest bias voltages shown, all three approaches give almost identical results, which means that in this case $N_{\text{B}}=6$ bath sites are sufficient. At intermediate voltages $\phi/\Gamma = 1/2$ and $1$ the ED results deteriorate, while CI remains comparable to MPS over a wide frequency range. This contrast between CI on the one hand and ED on the other can be attributed to the fits shown in the insets of Fig.~\ref{fig:GF_MPS_CI_compare}. While the six bath sites of ED provide an inaccurate fit, the eight bath sites of CI capture the physical hybridization function~\footnote{The Keldysh component of the hybridization function can be inferred from the fluctuation-dissipation theorem in Eq.~(\ref{eqn:fluc_diss}). Is is not shown, since it is less suited to illustrate the difference in the quality of the fit.} quite well. In equilibrium at $\phi=0$, CI keeps performing significantly better than ED. With $-\IIm[G^{\text{R}}(0)]\Gamma\approx 0.885$, CI is very close to the $T=0$ Friedel sum rule $-\IIm[G^{\text{R}}(0)]\Gamma=1$~\cite{lang.66,la.am.61}. It is thus also comparable to the result using MPS of $-\IIm[G^{\text{R}}(0)]\Gamma\approx 0.879$ and NRG with $-\IIm[G^{\text{R}}(0)]\Gamma\approx 0.891$ at this temperature.

Compared to CI, CASCI does not change these results as can be seen in Fig.~\ref{fig:GF_CI_CASCI_compare} within App.~\ref{subsec:CASCI_results}.

Since there are more powerful approaches such as NRG~\cite{wils.75, bu.co.08, zi.pr.09} available for equilibrium conditions, we will now investigate the temperature dependence of the conductance for the equilibrium impurity problem and compare it with NRG as implemented in~\cite{zitk.21}. The $\phi=0$ conductance is obtained by analytically differentiating the Meir-Wingreen formula~\cite{ha.ja, me.wi.92, jauh_link} with respect to the voltage $\phi$, which reduces for our leads in Eq.~\eqref{eqn:hyb_ph} (NRG) and in Eq.~\eqref{eqn:hyb_ph_real} (AMEA) to
\begin{align}
\mathcal{G}(\phi=0) &= \left.\frac{\partial j(T)}{\partial \phi}\right|_{\phi=0}\\ 
&= \int_{\mathrlap{-\infty}}^{\mathrlap{\infty}}\dd\omega\, \frac{\exp(\omega/T)}{\pi T [\exp(\omega/T) + 1]^2} \times \notag \\[-1mm]
&\hspace{2.4cm}\times\IIm[\Delta^{\text{R}}(\omega)] \IIm[G^{\text{R}}(\omega)]. \label{eqn:cond_MW_phi0}
\end{align}
\begin{figure}[t]
\centering
\includegraphics[width=0.9\linewidth]{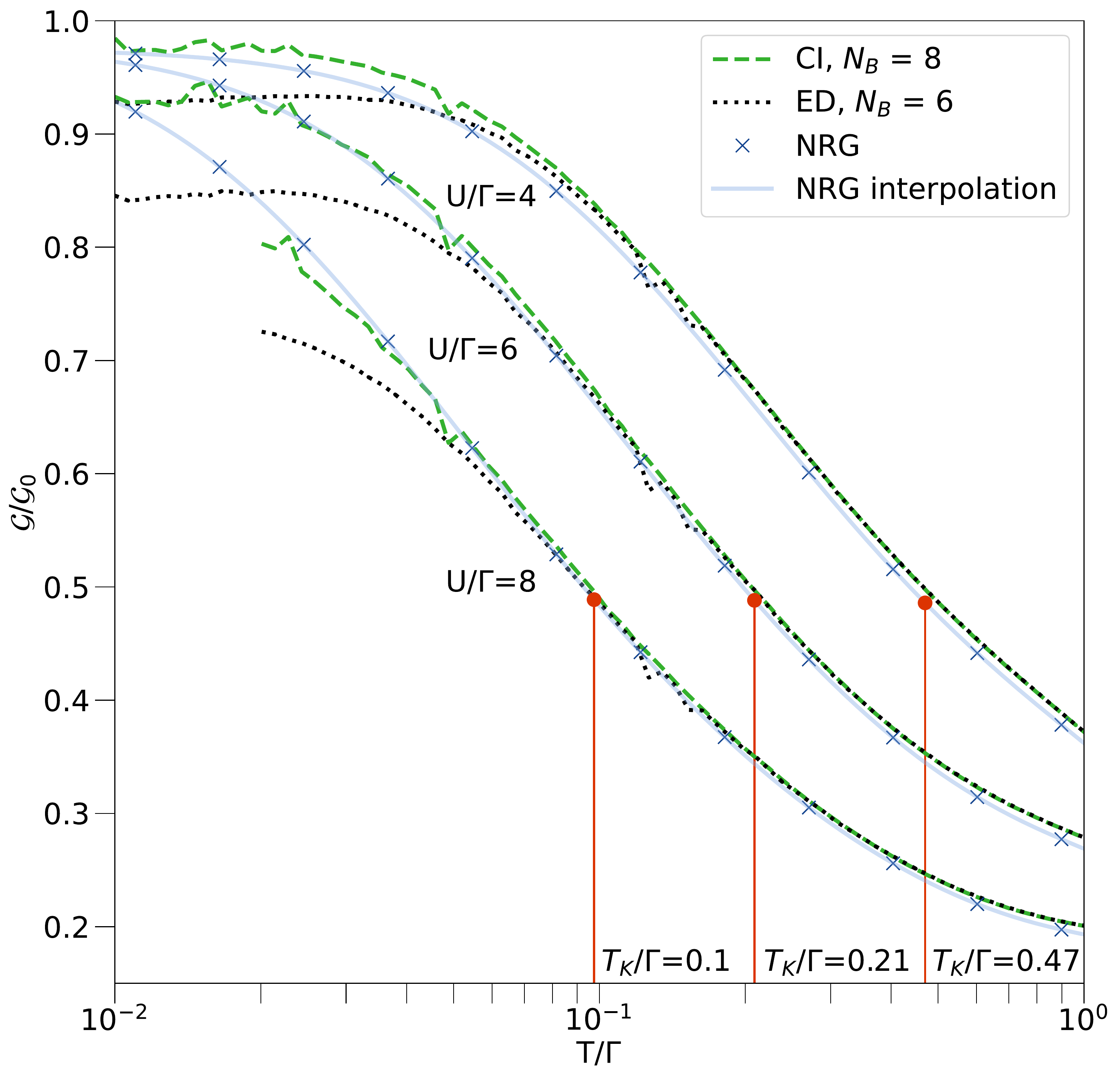}
\caption{Normalized equilibrium conductance $\mathcal{G}(\phi=0)/\mathcal{G}_{0}$ as a function of temperature for the Hubbard interaction $U/\Gamma=4, 6$ and $8$ with $N_{\text{B}}$ bath sites specified in the legend. Comparison between CI and ED as solvers for AMEA as well as NRG~\cite{zitk.21}.}
\label{fig:conductance_CI_NRG_compare}
\end{figure}
It is well known~\cite{bu.co.08,hews} that in the Anderson and Kondo model the conductance increases as the temperature decreases. Physically, this is due to the emergence of the Kondo peak in the spectral function. In the limit $T\to 0$, the spectral function at zero frequency is independent of the interaction strength, as stated by the Friedel sum rule~\cite{lang.66,la.am.61}. Consequently, following Eq.~\eqref{eqn:cond_MW_phi0}, the dc conductance reaches the same value $\mathcal{G}_{0}=1/\pi$ in our units for $T\to 0$, independent of the interaction.
\begin{figure}[t]
\centering
\includegraphics[width=0.9\linewidth]{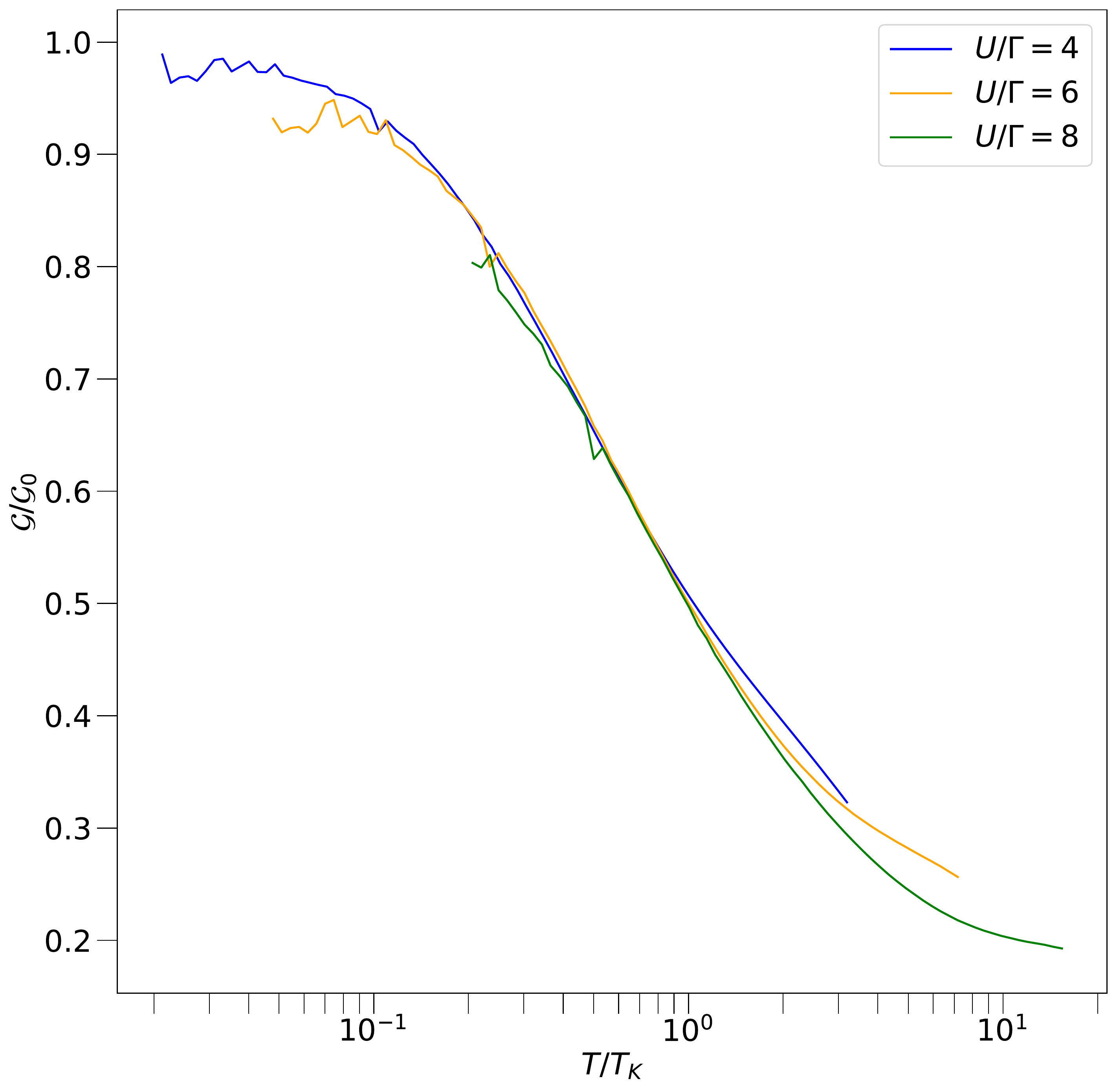}
\caption{Normalized equilibrium conductance $\mathcal{G}(\phi=0)/\mathcal{G}_{0}$ as a function of the scaled temperature $T/T_{\text{K}}$ for the Hubbard interactions $U/\Gamma=4, 6$ and $8$ obtained with CI using $N_{\text{B}}=8$.}
\label{fig:conductance_CI_scaled}
\end{figure}

Figure~\ref{fig:conductance_CI_NRG_compare} shows the equilibrium conductance $\mathcal{G}(\phi=0)$ obtained via CI and ED as solvers for AMEA, and NRG. For better orientation regarding the relative temperature scale, we show the Kondo temperature for each interaction strength $U$~\footnote{Following the Refs.~\cite{hews,bu.co.08,do.ga.15}, the Kondo temperature $T_{\text{K}}$ is obtained with NRG using the conductance $\mathcal{G}(T=0,\phi=0) = 2 \mathcal{G}(T=T_{\text{K}},\phi=0)$.}. At high temperatures all methods agree well, apart from a small offset for NRG. This behavior is probably due to NRG using the flat band spectrum in Eq.~\eqref{eqn:hyb_ph} instead of its smoothed version in Eq.~\eqref{eqn:hyb_ph_real}. For $U/\Gamma=4$, CI gives the quantitatively correct behavior of the conductance down to the smallest temperature considered $T/\Gamma\approx 0.01$. For $U/\Gamma=6$ and $8$ this is valid down to threshold temperatures~\footnote{The threshold temperatures provided mark the point where the conductance obtained with CI starts to deviate more than $2\%$ from the NRG reference.} $T/\Gamma\approx0.015$ and $0.023$, which are smaller than their respective $T_{\text{K}}$. In any case, CI is performing significantly better than ED and comparable to NRG for small temperatures. Comparing the conductance for different $U$, we note that with increasing $U$ the temperature at which CI starts to deviate from the NRG results increases. This is to be expected, because first, the HF reference state is not the best choice for large $U$~\cite{zg.gu.12, li.de.13}, and second, for an accurate description of larger $U$, one requires a larger state space than CI can provide. ED on the other hand starts to deviate from NRG always at the same temperature independent of $U$, since the fit of the hybridization function does not change as $U$ changes.

Compared to CI, CASCI does not change these results as can be seen in Fig.~\ref{fig:conductance_CI_CASCI_compare} within App.~\ref{subsec:CASCI_results}.

Figure~\ref{fig:conductance_CI_scaled} shows the equilibrium conductance $\mathcal{G}(\phi=0)$ obtained with CI as already discussed in Fig.~\ref{fig:conductance_CI_NRG_compare} but with the temperature rescaled in units of the Kondo temperature. As can be seen, the conductances for different interactions $U$ collapse quite well onto a single curve as expected and  observed for NRG~\cite{bu.co.08,ha.we.14}, at least up to $T\sim T_{\text{K}}$. In other words, the conductance for systems of different moderate to strong interactions is characterized by a single energy scale $T_{\text{K}}$.

In the following, we consider the system out of equilibrium under challenging conditions such as small temperatures and strong correlations. We investigate the current, conductance and spectral function as functions of the bias voltage at two temperatures $T=T_{\text{K}}/4$ and $T=2 T_{\text{K}}$ below and above the Kondo temperature. For the smaller temperature, we expect the presence of a Kondo peak, and thus an enhanced conductance~\cite{bu.co.08}. As the voltage increases, we expect the Kondo peak to split~\cite{me.wi.93,fr.ha.02,le.sc.05,sh.ro.06,fr.ke.10,kr.kl.19,ha.he.07,ande.08,do.ga.15,fu.do.18,fu.ba.20,er.gu.22u}, in turn reducing the dc conductance. 

\begin{figure}[t]
\centering
\includegraphics[width=0.95\linewidth]{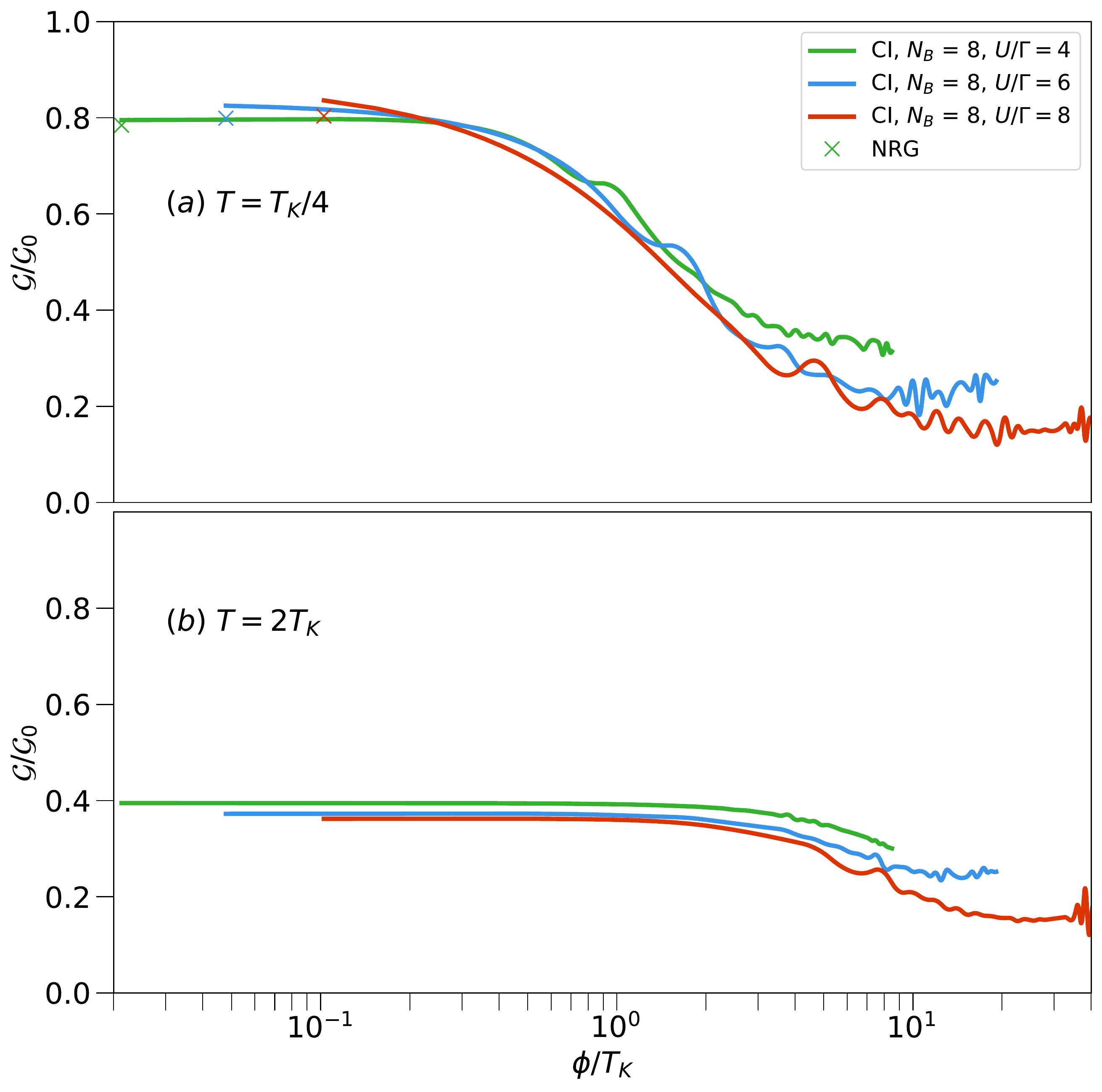}
\caption{Normalized conductance $\mathcal{G}/\mathcal{G}_{0}$ as function of the scaled voltage $\phi/T_{\text{K}}$ for $N_{\text{B}}$ bath sites specified in the legend, two different temperatures (a) $T=T_{\text{K}}/4$, (b) $2T_{\text{K}}$ and different Hubbard interaction $U/\Gamma=4, 6$ and $8$ using the CI solver for AMEA ($\phi\geq 0$). The crosses denote equilibrium ($\phi=0$) results obtained with NRG.}
\label{fig:current_CI}
\end{figure}

\begin{figure}[t]
\centering
\includegraphics[width=0.95\linewidth]{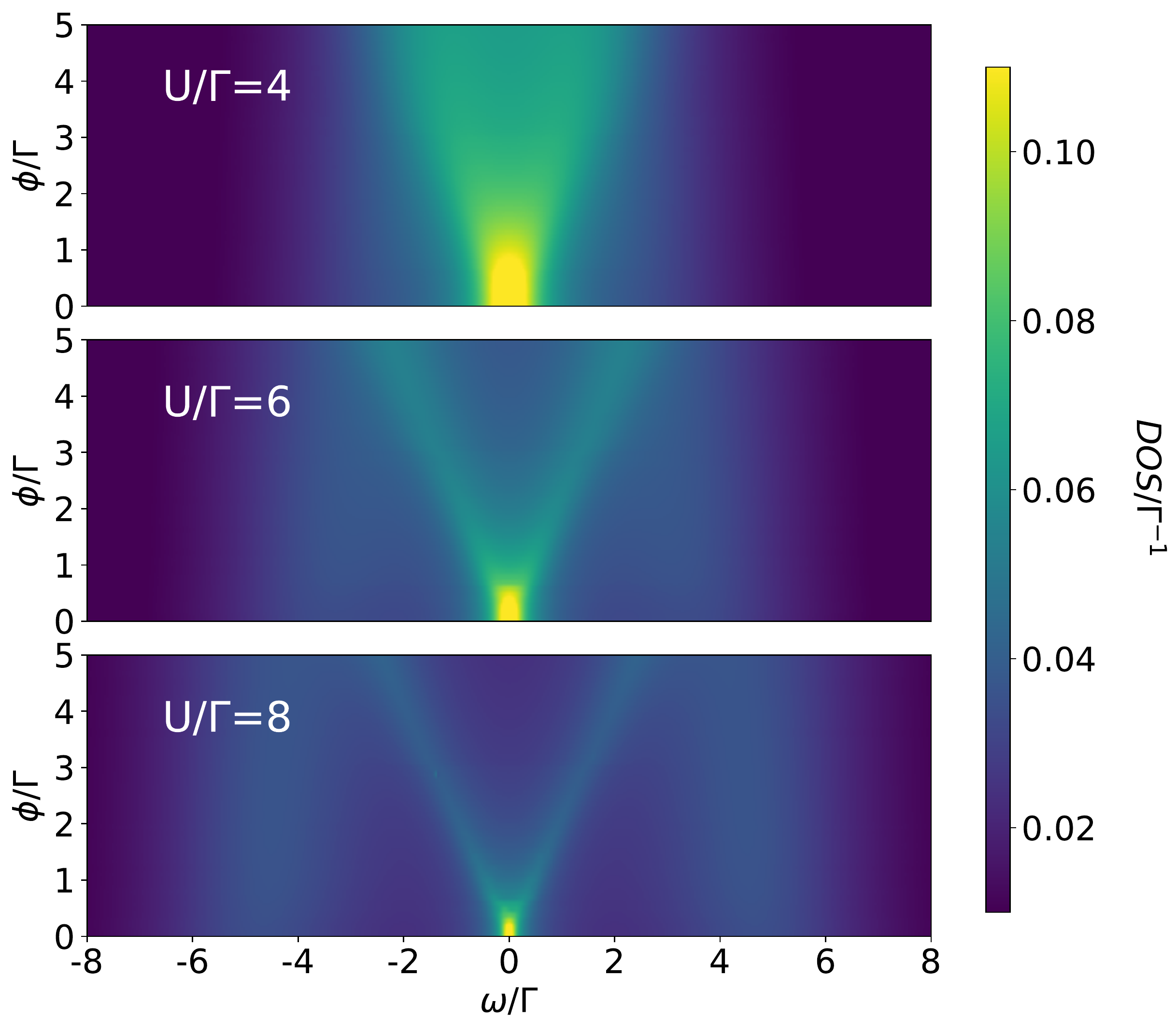}
\caption{Spectral function obtained via the CI solver for AMEA as a function of $\omega$ and $\phi$ for different $U/\Gamma=4, 6$ and $8$ at $T=T_{\text{K}}/4$ and $N_{\text{B}}=8$ bath sites.}
\label{fig:Kondo_split}
\end{figure}

Figure~\ref{fig:current_CI} shows the conductance~\footnote{The conductance is obtained as the analytical derivative of splines used to interpolate the current. The oscillatory behavior in the conductance is an artifact of this procedure. We use \texttt{sci\-py.in\-ter\-po\-late.Uni\-vari\-ate\-Spline}, with the parameters $s=0$ to ensure that the raw data points are interpolated, and $k=3$ for cubic splines.}, while Fig.~\ref{fig:Kondo_split} displays the spectral function for various Hubbard interactions.
For comparison, we also show the NRG conductance in equilibrium at $\phi = 0$, which agrees quantitatively with the results of CI for the different $U$ considered. Besides the enhanced conductance due to the Kondo peak at the smaller temperature, we observe that for $T=T_{\text{K}}/4$, the conductances for the different interactions $U/\Gamma=4, 6$ and $8$ exhibit the same scaling behavior with $\phi/T_{\text{K}}$~\cite{bu.co.08}. For $T=2T_{\text{K}}$, corrections to scaling due to the finite bandwidth~\cite{ha.we.14} can be already observed.
%
%

Shifting the focus onto the spectral functions in Fig.~\ref{fig:Kondo_split}, we observe that with larger interaction strength, the Kondo peak becomes sharper and splits at smaller voltage which leads to a more rapid decline in the conductance of Fig.~\ref{fig:current_CI}.

\section{\label{sec:conclusion}Conclusion}
In this paper, we used CI and CASCI~\cite{he.jo, sh.sc.99, zg.gu.12, li.de.13} to solve the auxiliary master equation for interacting systems in AMEA~\cite{ar.kn.13,do.nu.14}. On the one hand, this allowed us to treat larger auxiliary systems with more optimization parameters in the mapping procedure as compared to an ED based solver~\cite{do.nu.14} and about as many optimization parameters as a MPS based solver~\cite{do.ga.15} provides. This is key, since the accuracy of AMEA increases exponentially with the number of optimization parameters~\cite{do.so.17}. On the other hand, CI and CASCI require smaller runtimes than the MPS based solver~\cite{do.ga.15}. Using CI and CASCI as solvers for AMEA is furthermore numerically much cheaper than methods like quantum quasi Monte Carlo~\cite{be.ba.21}, fork tensor network approaches~\cite{ba.zi.17} or inchworm QMC~\cite{er.gu.22u} which have been used to address long times or the nonequilibrium steady state.

To illustrate the improvement of using CI and CASCI, we investigated the steady state properties of the Anderson impurity model as functions of the bias voltage $\phi$ and the temperature $T$ well below the Kondo temperature $T_{\text{K}}$. We showed that CI and CASCI provide good quality spectral quantities comparable to MPS, and an equilibrium conductance comparable to NRG starting above and going below the Kondo temperature for large interactions. More specifically, we assessed the quality of the spectral functions of CI and CASCI by comparing them with those of MPS and NRG via the $T=0$ Friedel sum rule $-\IIm[G^{\text{R}}(0)]\Gamma=1$~\cite{lang.66,la.am.61}. At the temperature $T/\Gamma=0.05$, CI and CASCI ($\approx0.885$) perform slightly better than MPS ($\approx0.879$) and slightly worse than NRG ($\approx0.891$). Comparing the equilibrium conductance with NRG allowed us to infer the lowest temperatures we can reach reliably for a range of interactions. For the largest interaction $U/\Gamma=8$ considered we obtain the largest threshold temperature $T/\Gamma\approx0.023=0.23T_{\text{K}}/\Gamma$. The computed equilibrium and nonequilibrium conductance quite accurately collapses into a single function of $T/T_{\text{K}}$ and $\phi/T_{\text{K}}$  for different values of the interactions $U/\Gamma=4, 6$ and $8$, as long as the energies are much smaller (about one-tenth) than the bandwidth. The low-temperature spectrum displays a Kondo peak that splits up with increasing bias voltage.

Having illustrated the accuracy of CI and CASCI, we now comment on their computational costs. The CI solver with $N_{\text{B}}=6$ bath sites and a wall clock time of ${\sim}\unit[5]{min}$ takes comparably long as ED with $N_{\text{B}}=4$ (${\sim}\unit[3]{min}$). CI with $N_{\text{B}}=8$ bath sites requires ${\sim}\unit[45]{min}$ which can be compared with ED with $N_{\text{B}}=6$ (${\sim}\unit[30]{min}$). CASCI takes about $1.5\text{--}2$ times as long as CI, which is still quite good~\cite{Note12}. These short times make CI/AMEA an appealing impurity solver for nonequilibrium DMFT. With respect to a plain ED/AMEA solver, one can achieve the same accuracy  with an order of magnitude smaller wall clock time per DMFT iteration.

Up to now, we have not taken advantage of the full potential of CI, because so far we have only used the HF basis. For larger electron-electron interactions, it is probably more advantageous to use the natural orbital basis~\cite{zg.gu.12, li.de.13}. We expect its implementation to further improve the accuracy of the impurity solver. Another avenue to improve CI consists in constructing the many-body basis via adaptive sampling~\cite{me.tu.19,tu.le.16.da,tu.fr.20}.

\section*{Acknowledgements}
This work was supported by the Austrian Science Fund (FWF) within Project P 33165-N, as well as NaWi Graz. The computational results presented have been obtained using the D-Cluster Graz. We use the QuSpin library~\cite{wein.22} to set up the basis of many-body states and express the interacting Lindbladian with it. For the reference computations in equilibrium, we use NRG Ljubljana~\cite{zitk.21}.

\appendix

\section{\label{sec:CASCI}Complete active space extension}
In equilibrium, CASCI poses a straightforward way to improve CI by including more excitations. Therefore, we present its realization as a solver for AMEA and the results obtained using it. However, in our case, CASCI does not seem to introduce a significant improvement.

\subsection{\label{subsec:CASCI_method}Method}
\begin{figure}[t]
\centering
\includegraphics[width=\linewidth]{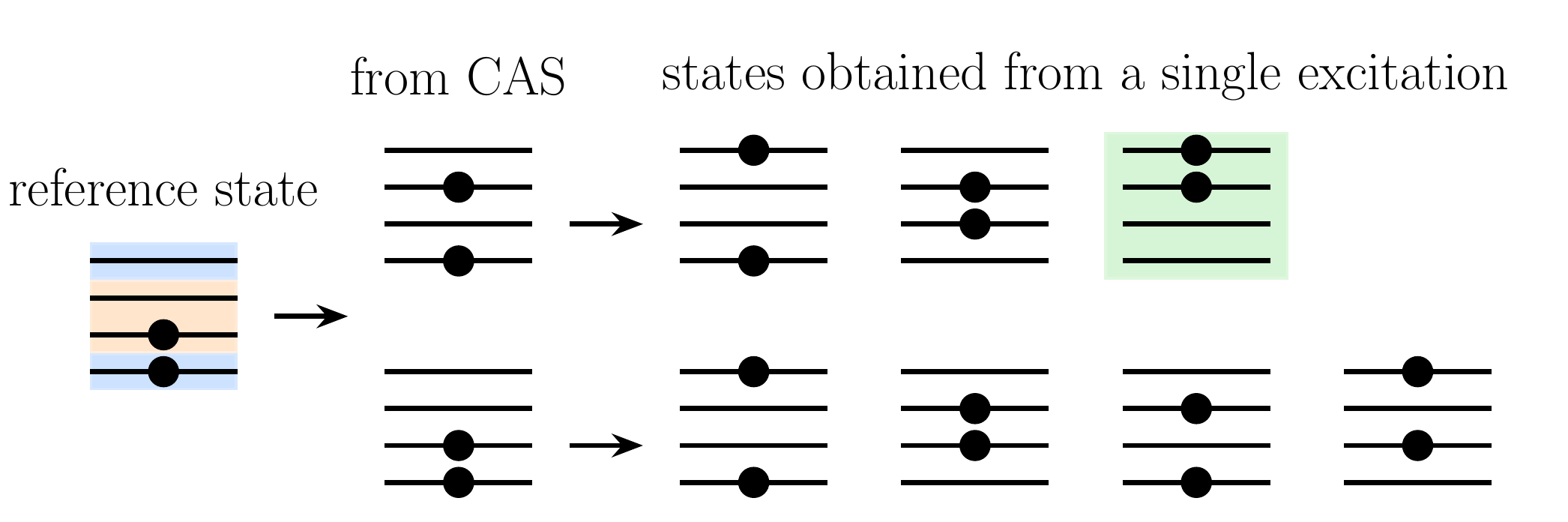}
\caption{Construction of the many-body basis in CASCI using the orbitals labeled by normal modes. On the left is the reference state. The two orbitals highlighted in orange form the active space. First, we create all states reached by applying single PH excitations in the active space~\cite{zg.gu.12}. Then, each of the obtained states undergoes the procedure depicted in Fig.~\ref{fig:CI}. States that appear multiple times are discarded. We obtain an additional state compared to CI, which is highlighted in green. Another possible choice of active space consists of the two orbitals highlighted in blue.}
\label{fig:CAS}
\end{figure}

In equilibrium, starting only from a single reference state may become insufficient once many low-lying states are close to the Hartree-Fock ground state. This can be improved by using the complete active space (CAS) extension of CI. Therein one takes into account further states, which are nearly degenerate with the reference state in terms of the effective noninteracting Hamiltonian. The basis then consists of the reference state, the nearly degenerate states as well as PH excitations applied to them. To produce the nearly degenerate states from the reference state, one selects a set of nearly degenerate, \textquoteleft active\textquoteright\, orbitals (active space) on which a fixed number of particles is distributed in all distinct combinations~\cite{he.jo, zg.gu.12, li.de.13}. In Fig.~\ref{fig:CAS}, the first two columns illustrate this for a single particle on two active orbitals. Further basis states generated by single PH excitations are shown in the remaining columns.

In nonequilibrium, the reference state is the steady state and the orbitals are labeled by normal modes. Here, we restrict ourselves to active spaces consisting of two orbitals and their spin-flipped counterparts. Following the notion of the steady state resembling a Fermi sea, we will show in Sec.~\ref{subsec:CASCI_results}, that the active space is best selected by employing orbitals of smallest $|\IIm(\varepsilon_{i})|$. Starting from the single reference state, we will generate the \textquoteleft nearly degenerate\textquoteright\, states by applying single PH excitations to the orbitals within the active space. From the reference state and the \textquoteleft nearly degenerate\textquoteright\, states, the remaining basis states are generated via PH excitations.

In practice, we follow the procedure outlined in Sec.~\ref{sec:CI} with the following additional steps. For the steady state, before requiring $N_{\sigma}-\tilde{N}_{\sigma}=0$, we consider in addition the states generated with eight $\Bar{P}$-operators, i.e. four PH excitations, where the additional PH excitation shifts particles within the active space.
For the Green's function, we similarly apply nine $\Bar{P}$-operators.

\subsection{\label{subsec:CASCI_results}Results}
\begin{figure}[t]
\centering
\includegraphics[width=0.9\linewidth]{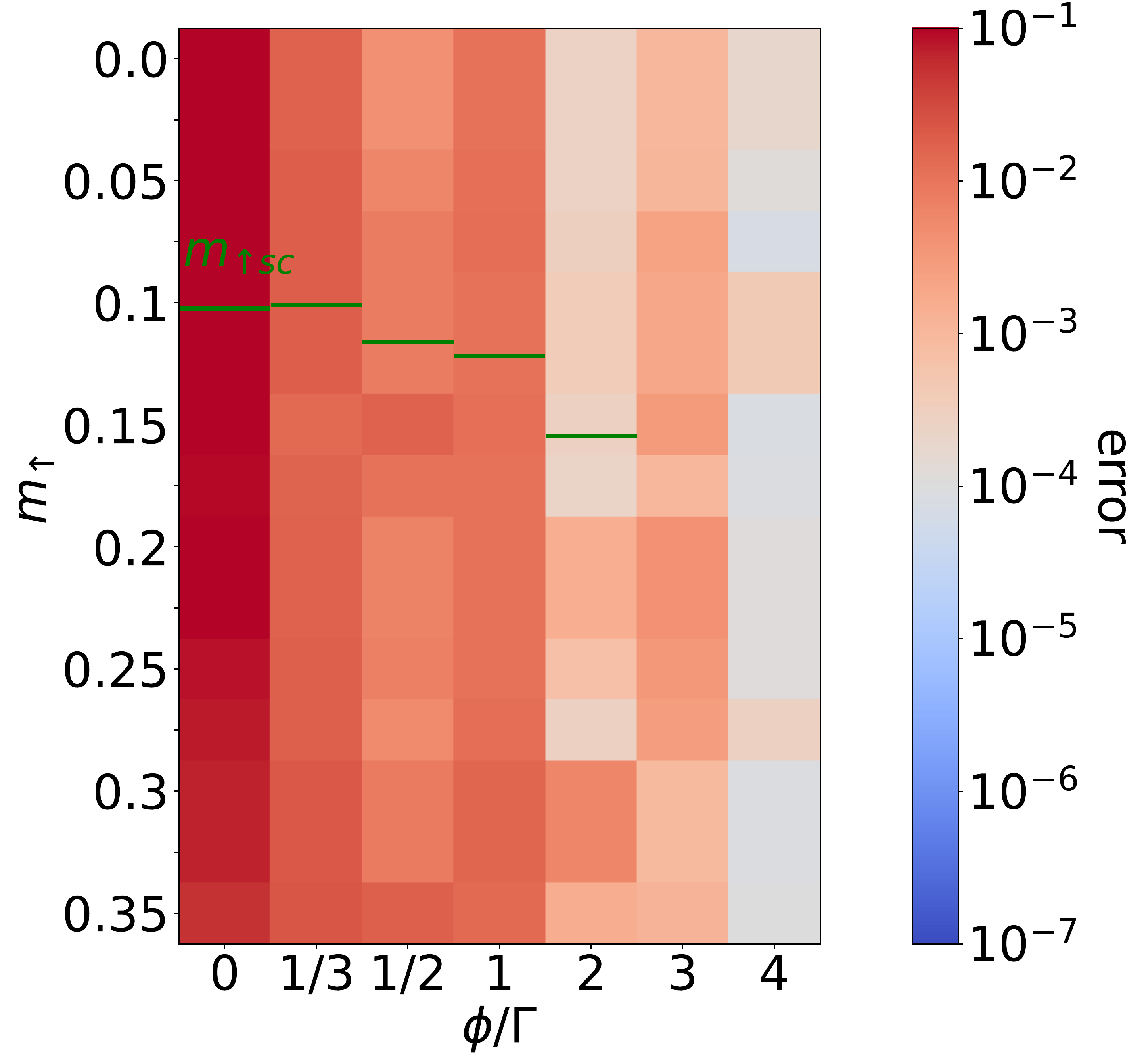}
\caption{Matrix plot of the error defined in Eq.~\eqref{eqn:error_calc} for CASCI as solver for AMEA given different voltages $\phi$ and parameters $m_{\uparrow}$ with resolution $\Delta m_{\uparrow}=0.025$. For $m_{\uparrow}\in[0.375,0.5]$, the error jumps two orders of magnitude for moderate and large voltages and is omitted here. Green lines denote self-consistently determined parameters $m_{\uparrow\text{sc}}$~\cite{Note9}. The remaining parameters are the temperature $T/\Gamma=0.05$, Hubbard interaction $U/\Gamma=6$, and $N_{\text{B}}=6$ bath sites.}
\label{fig:CASCI_error_colorplot}
\end{figure}

First, we check whether the choice for the parameter $m_{\uparrow}$ used for CI in Sec.~\ref{sec:dof} is convenient for CASCI as well. It is required to fully determine the interacting Lindbladian given in Eq.~\eqref{eqn:Lindblad_int}. The system parameters are the same as in Sec.~\ref{sec:dof}, low temperature $T/\Gamma=0.05$, large interaction $U/\Gamma=6$ and $N_{\text{B}}=6$ bath sites restriced by the reference ED. Figure~\ref{fig:CASCI_error_colorplot} shows the difference between CASCI and ED, calculated via Eq.~\eqref{eqn:error_calc}. 
As for CI, we omit the worst and partially non-convergent results obtained for $m_{\uparrow}>0.35$ around half-filling $m_{\uparrow}=0.5$. Even though the self-consistently determined values $m_{\uparrow\text{sc}}$~\cite{Note9} perform better than half-filling for small to intermediate voltages, the minimal error for both methods is located around $m_{\uparrow} = 0.3$.
\begin{figure}[b]
\centering
\includegraphics[width=0.85\linewidth]{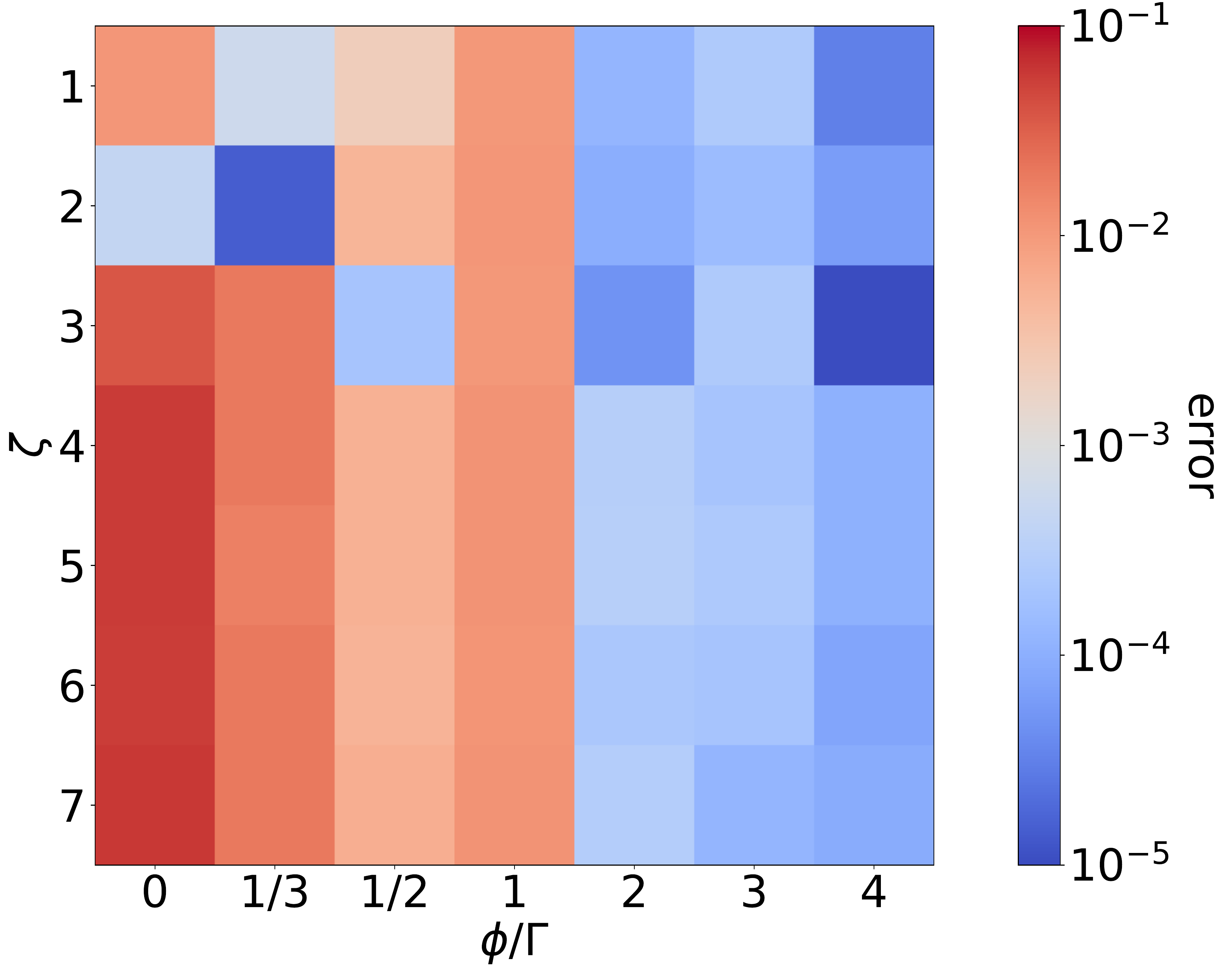}
\caption{Matrix plot of the error defined in Eq.~\eqref{eqn:error_calc} for CASCI as solver for AMEA given different voltages $\phi$ and using different active spaces enumerated by $\zeta$. Details on the enumeration can be found in the main text. The remaining parameters are the temperature $T/\Gamma=0.05$, Hubbard interaction $U/\Gamma=6$, and $N_{\text{B}}=6$ bath sites.}
\label{fig:CASCI_intend_error_colorplot}
\end{figure}

The second method parameter for CASCI is the composition of the active space. In equilibrium, the active space can be chosen symmetrically around the highest occupied orbital~\cite{zg.gu.12,li.de.13} similar to the illustration in Fig.~\ref{fig:CAS}. Since the eigenvalues of the Lindbladian are not energies, but complex values, we need another way to measure how \textquoteleft close\textquoteright\, states are to each other.

As established in Sec.~\ref{sec:CI}, the imaginary part of the eigenvalues $\IIm(\varepsilon_{i})$ takes the role of the single-particle energies separating the occupied and empty orbitals. In analogy to Eq.~\eqref{eqn:time_evolution}, the time-evolution of the state obtained from a PH excitation applied to the noninteracting steady state is
\begin{align} \label{eqn:time_evolution_ph}
\ee^{L_0 t}\Bar{\xi}_{i}\xi_{j} \ket{\rho_{\infty 0}} &= \ee^{L_0 t}\Bar{\xi}_{i}\ee^{-L_0 t}\ee^{L_0 t}\xi_{j}\ee^{-L_0 t} \ket{\rho_{\infty 0}} \notag\\
&= \ee^{\ii(\varepsilon_{j}-\varepsilon_{i}) t} \Bar{\xi}_{i}\xi_{j} \ket{\rho_{\infty 0}}.
\end{align}
A smaller difference in $\IIm(\varepsilon_{j}-\varepsilon_{i})$ thus implies a longer-lived state. Starting from $t\to\infty$ and going to shorter times, the states contributing to the long-time behavior can be sorted by $\IIm(\varepsilon_{j}-\varepsilon_{i})$ starting from the steady state. It turned out in practice that each orbital and its spin-flipped counterpart share the same imaginary part of the single-particle energies $\IIm(\varepsilon_{i\uparrow})=\IIm(\varepsilon_{i\downarrow})$. Therefore we restrict ourselves here to states consisting of one PH excitation between orbitals with $\IIm(\varepsilon_{i})=-\IIm(\varepsilon_{j})$ plus their respective spin-flipped counterparts. This makes $\IIm(\varepsilon_{i})$ a sufficient distance measure.
\begin{figure}[t]
\centering
\includegraphics[width=\linewidth]{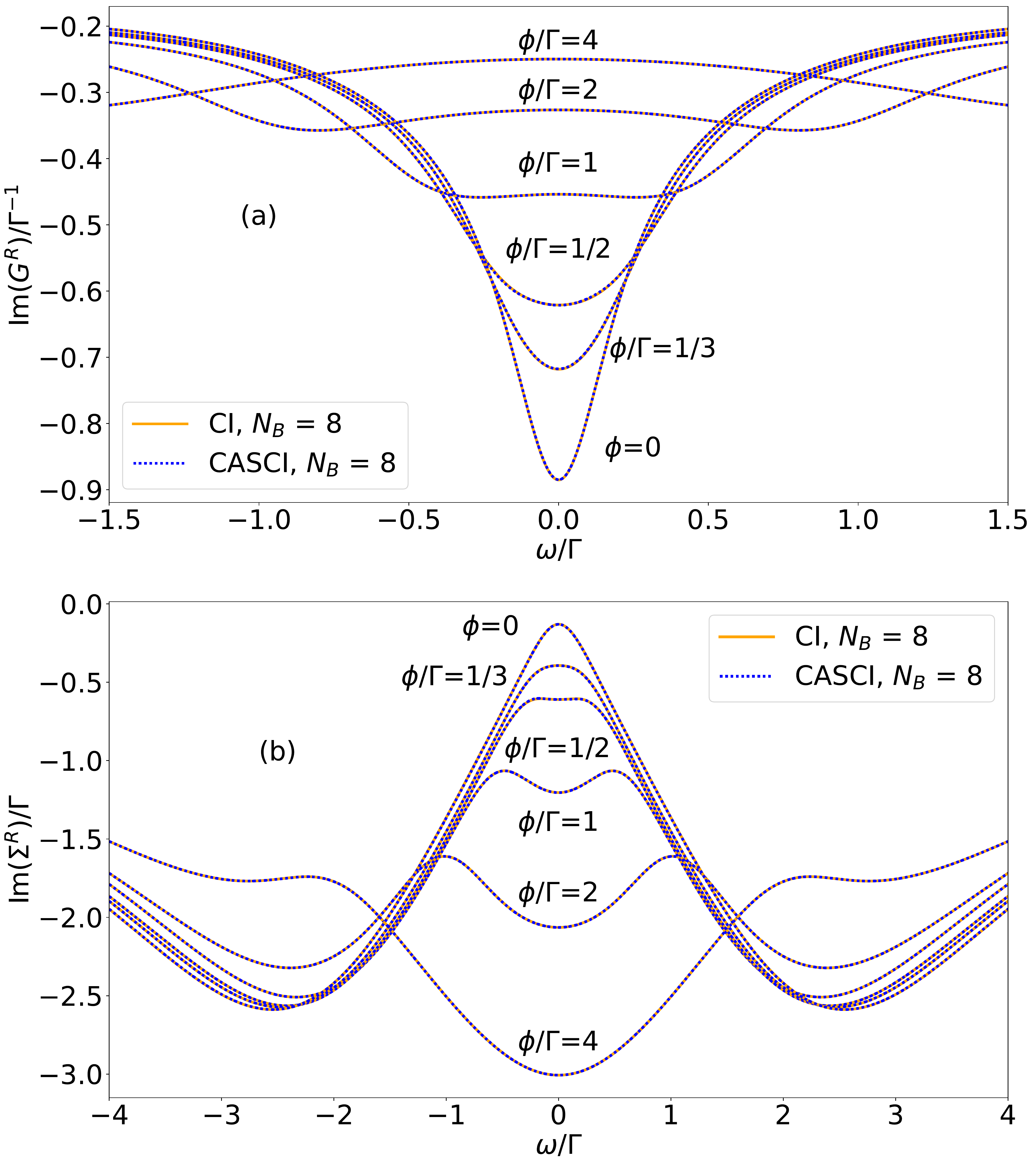}
\caption{Comparison of the imaginary parts of the retarded components of the (a)~Green's function $G$ and the (b)~selfenergy $\Sigma$ for temperature $T/\Gamma=0.05 $, Hubbard interaction $U/\Gamma=6$, $N_{\text{B}}$ bath sites specified in the legend and various voltages $\phi$ obtained from CI and CASCI as solvers for AMEA.}
\label{fig:GF_CI_CASCI_compare}
\end{figure}

To clarify whether $\IIm(\varepsilon_{i})$ is a reliable criterion to select the active space, we perform a parameter sweep, which is shown in Fig.~\ref{fig:CASCI_intend_error_colorplot}. The quantity $\zeta$ therein refers to the active spaces sorted by their distance from the steady state measured by $\IIm(\varepsilon_{i})$. In terms of Fig.~\ref{fig:CAS}, $\zeta=1$ refers to the orange active space, while $\zeta=2$ corresponds to the active space formed by the orbitals highlighted in blue. Higher $\zeta$ refer to more distant orbitals which are chosen symmetrically around the boundary between occupied and empty orbitals. Each of these active spaces consists of up to four orbitals, two occupied and two unoccupied ones with opposite spin in terms of the $\xi$-operators. In agreement with our expectations, Fig.~\ref{fig:CASCI_intend_error_colorplot} shows that, given our parameters, the active spaces labeled by $\zeta=1$ and $2$ with the smallest $\IIm(\varepsilon_{i})$ yield the best result. Hence throughout this appendix, we will choose the active space based on the smallest $\IIm(\varepsilon_{i})$.

Figure~\ref{fig:GF_CI_CASCI_compare} shows a comparison of $\IIm(G^\text{R})$ and $\IIm(\Sigma^\text{R})$ for $T/\Gamma=0.05$, $U/\Gamma=6$, $N_{\text{B}}$ specified in the legend and various voltages $\phi$ which are computed using CI and CASCI as solvers for AMEA. It is evident, that CASCI and CI give the same results. Thus CASCI provides results which are significantly better than ED and on par with MPS for all considered bias voltages. The $T=0$ Friedel sum rule $-\IIm[G^{\text{R}}(0)]\Gamma=1$~\cite{lang.66,la.am.61} allows to quantify how well CASCI with $-\IIm[G^{\text{R}}(0)]\Gamma\approx 0.885$ performs in comparison to MPS with $-\IIm[G^{\text{R}}(0)]\Gamma\approx 0.879$ and NRG with $-\IIm[G^{\text{R}}(0)]\Gamma\approx 0.891$.
\begin{figure}[t]
\centering
\includegraphics[width=0.9\linewidth]{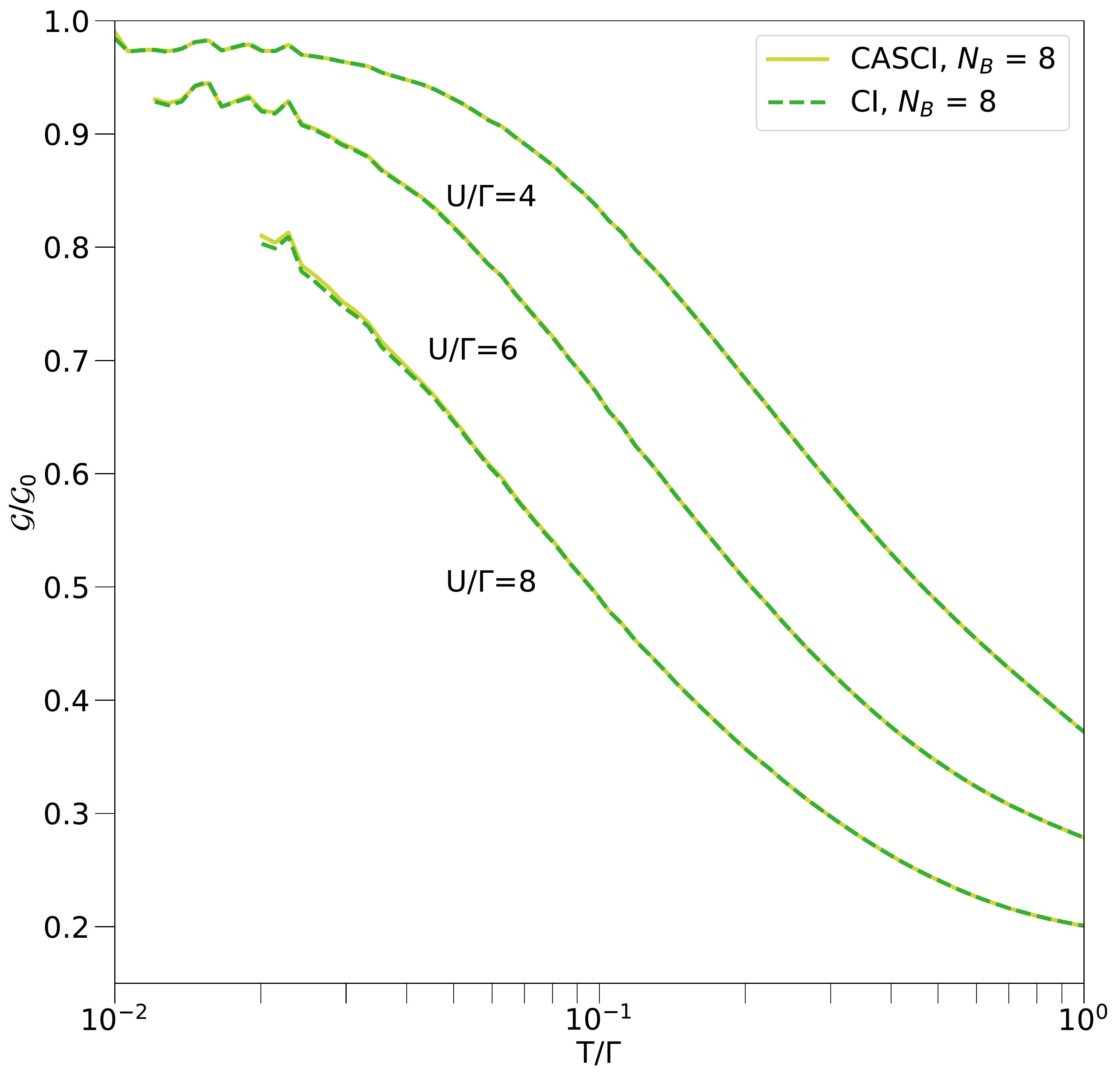}
\caption{Normalized equilibrium conductance $\mathcal{G}(\phi=0)/\mathcal{G}_{0}$ as a function of temperature for the Hubbard interaction $U/\Gamma=4, 6$ and $8$ with $N_{\text{B}}$ bath sites specified in the legend. Comparison between CI and CASCI as solvers for AMEA.
}
\label{fig:conductance_CI_CASCI_compare}
\end{figure}

Figure~\ref{fig:conductance_CI_CASCI_compare} shows the equilibrium conductance $\mathcal{G}(\phi=0)$ obtained using Eq.~\eqref{eqn:cond_MW_phi0} via CI and CASCI as solvers for AMEA. Also here, CASCI and CI give the same results. Hence CASCI provides the quantitatively correct behavior of the conductance down to the smallest temperature considered $T/\Gamma\approx 0.01$ for $U/\Gamma=4$ and down to the threshold temperatures $T/\Gamma\approx0.015$ and $0.023$ for $U/\Gamma=6$ and $8$, which are smaller than their respective $T_{\text{K}}$.

\bibliographystyle{prsty}
\bibliography{references_database,my_refs}

\end{document}